\documentclass[aps,epsfig,nofootinbib]{revtex4}
\pdfoutput=1
\usepackage{graphicx}
\usepackage{dcolumn}
\usepackage{amsmath}
\usepackage{amssymb}
\usepackage{epsfig}
\usepackage[colorlinks,linkcolor=blue,citecolor=blue,urlcolor=blue]{hyperref}

\begin{document}

\title{Preinflationary dynamics in loop quantum cosmology: Monodromy Potential}
\author{Manabendra Sharma$^1$ \footnote{E-mail address: manabendra@zjut.edu.cn}}
\author{M. Shahalam$^1$ \footnote{ E-mail address: shahalam@zjut.edu.cn}}
\author{Qiang Wu$^1$ \footnote{ E-mail address: wuq@zjut.edu.cn}}
\author{Anzhong Wang$^{1,2}$ \footnote{E-mail address: Anzhong$\_$Wang@baylor.edu; Corresponding Author }}
\affiliation{$^{1}$Institute for Advanced Physics $\&$ Mathematics,
Zhejiang University of Technology, Hangzhou, 310023, China\\
$^2$GCAP-CASPER, Department of Physics, Baylor University, Waco, TX, 76798-7316, USA }

\date{\today}

\begin{abstract}
In this article we explore the pre-inflationary background dynamics of an FLRW universe sourced by a scalar field with monodromy potential in LQC framework. In particular we calculate the number of e-folds, $N_{inf}$, produced during the slowly rolling phase of the inflation and find out the critical value of the ratio of the kinetic to potential energy, $r_w^c$, at the quantum bounce that is required to produce $N_{inf}\simeq 60.$ Two different monodromy potentials, namely, linear and quadratic with a modulation term are investigated to this effect. The effects on the value of $N_{inf}$ due to parameters associated with the strength, decay constant and the phase factor of the modulation term are calculated. In addition to this we present the qualitative picture of the background dynamics by carrying out a dynamical system analysis. We produce the phase portraits and carry out a detailed linear stability analysis of the finite fixed points, if any, for each of the potentials. 
\end{abstract}

\maketitle
 
\section{Introduction}\label{Introduction}
The inflationary cosmology as proposed by Alan Guth in the year 1981 \cite{Guth} gives a solution to some of the puzzles of the Standard Big Bang Cosmology by introducing an epoch of nearly exponential expansion of the early universe. Other than serving as a mechanism to solve problems like horizon, flatness and entropy etc., it is much more powerful in predicting the cosmos we see today \cite{Realized}. It provides a first ever causal explanation for the origin of inhomogeneities in the universe. To be precise, it predicts the primordial power spectra whose evolutions explain both the formation of large scale structure and small inhomogeneities present in the cosmic microwave background (CMB) \cite{Guth}\cite{CMB}. Today, the unprecedented success of the inflationary paradigm is based on observational precisions \cite{Success1}-\cite{Success3}. Though very successful, however, the paradigm also suffers from  issues  like initial singularity, trans-Planckian problem to give some instances. For a review on conceptual issues of inflationary cosmology see \cite{UC}.

For every expanding FLRW solution of GR there is a big-bang singularity if matter satisfies the standard energy conditions. This strong curvature singularity at which physics comes to halt is an artifact of the reason that we are pushing the classical theory of gravity to a region where it is no longer valid. Mathematically speaking, the geodesics are past incomplete and hence the affine parameter can not be extended to infinite past leading to a focusing of congruence of geodesics\cite{Raychaudhuri}\cite{LargeScaleStructure}. This causes the breakdown of the notion of space-time itself. The result of focusing theorem is valid both for null and time like geodesic\cite{EricPoisson}. It can be shown that this issue is independent of the symmetry of the metric considered unless and until certain energy conditions are hold good and depends solely on the average value of the Hubble parameter\cite{MyThesis1}. Neither Standard model of cosmology nor inflationary paradigm have a say on this issue. An alternative scenario, called Bounce, within the framework of classical theory of gravity requires either the consideration of nontrivial stress-energy tensor or the modification of the gravity sector of the Einstein-Hilbert action which is evident from the Raychaudhuri Equation in the absence of the centrifugal term \cite{Raychaudhuri}. In addition to solving the puzzles of standard model of cosmology \cite{SolutionHPFP}, these classical bouncing scenario involves circumvention of the singularity by assuming the universe to start from a contracting and then bouncing back to the expanding phase well before reaching the Planck length \cite{MyThesis1}. However these scenarios are also faced with serious setbacks like instability problem \cite{Instabilities}. For a review on classical bouncing scenario and their drawbacks see \cite{ReviewOnBouncing1},\cite{ReviewOnBouncing2}. Amongst all the drawbacks the fact that the universe becomes extremely dense early on suggests that quantum gravity effects would become important at such small length and one must adopt quantum gravity to explore the ``zone of ignorance." Therefore, it is natural to speculate that in a true theory of quantum gravity quantum mechanics would intervene to avert the singularity. 

Loop quantum gravity (LQG) is a candidate of quantum gravity theory where it takes the premise of gravity as a manifestation of geometry of space-time seriously and then systematically constructs a specific theory of quantum Riemanian geometry with rigour. For a review we refer the reader to  \cite{Thiemann1}. LQG stands out as a leading non-perturbative background independent approach to quantize gravity\cite{QuantumGravity1}. At its depth, this theory brings out a fundamental discreteness at Planck scale wherein the underlying geometric observables-such as areas of physical surfaces and volumes of physical regions are discrete in nature \cite{Asthekar1}\cite{Asthekar2}\cite{Bianchi1}. The avenue of loop quantum cosmology (LQC) is an application of LQG techniques to the symmetry reduced space-time, homogeneous space-time in particular\cite{Bojowald1}. In LQC the singularity is resolved in the sense that physical macroscopic observables such as energy density and curvature which diverge at the big bang in GR, have a finite upper bound. This finite of the macroscopic parameters above owes to the fact that they have a dependence on the microscopic parameter of the theory called the fundamental area gap whose smallest eigenvalue is nonzero \cite{Asthekar3}\cite{Asthekar4}\cite{Asthekar5}. Thus, a contracting FLRW universe would bounce back to an expanding one avoiding the formulation of singularity in LQC as there is a maximum value of energy density. This is achieved without adding any nontrivial piece of matter unlike in the case of classical bounce. This quantum bounce occurs purely due to quantum geometric effects which acts as a novel repulsive force that can be easily seen from the quantum corrected Friedmann and Raychaudhuri Equation to be discussed in Sec.\ref{Cosmology}. Also, it is to be noted that in all the different classes of space-times permitting different set of symmetries \cite{WE}, including Bianchi and Gowdy models, the singularity is resolved in the framework of LQC\cite{Asthekar6}-\cite{Brizuela2}. Based on the loop quantization of Brans-Dick theory the dynamics of loop quantum Brans-Dick cosmology has also been explored \cite{Jin}. For a review of singularity resolution in LQC see \cite{Asthekar9} and \cite{Singh3}.

In addition to the above, the fact that the universe must have expanded at least 50 e-folds so as to be consistent with the current observation leads to a problematic situation if the universe had expanded a little more than 70 e-folds. In fact, had it been so (which is true for a large class of inflationary models \cite{5}), it turns out that the wavelengths of all fluctuation modes which are currently inside the Hubble radius were smaller than the Planck length at the onset of the epoch of inflation. This is coined as trans-Planckian issue in \cite{6} which challenges the validity of the assumption that matter fields are quantum in nature but spacetime can still be treated classically which are used at the beginning of inflation in order to make predictions \cite{CMB}. Thus, once again it calls for quantum treatment of space-time. Moreover, the inflationary paradigm usually sets the BD vacuum state at the time when the wavelengths of fluctuations were well within the Hubble horizon during the phase of inflation. This treatment, however allows the ignorance of the dynamics prior to the onset of inflation, even when the modes were well inside the Hubble horizon. For more details regarding the sensitivity of the inflationary dynamics we refer the readers to \cite{6}\cite{8}. 

As cited above all these issues motivate to look for quantum gravity candidates and LQC stands out as, with robustness, a competing one that replaces the singularity by a quantum bounce which is followed by a desired slow roll inflation \cite{18}. Now LQC is in a position to undergo experimental tests and to look for observational signature of quantum bounce, pre-inflationary dynamics in current/forth-coming observations. In fact three major streams of calculations of cosmological perturbations, namely, \textit{dressed metric}, \textit{deformed algebra} and \textit{hybrid approaches}
 \cite{Asthekar9}, \cite{14}, \cite{Mendez} have been carried out and studied numerically and analytically in \cite{10},\cite{11},\cite{Zhu01},\cite{Zhu02},\cite{Wu}. It has been reported that the \textit{deformed algebra} approach is already inconsistent with current observations \cite{21}.

In this paper we investigate the dynamics of a universe sourced by scalar field with monodromy potential \cite{MonodromyPotential1} in an FLRW background in LQC framework. For readers interested in the origin and cosmological implication of monodromy potential, we refer them to \cite{MonodromyPotential}. We consider two cases of potentials: one in which the monodromy term is linear and the other with a quadratic term. This paper  is organized into two major sections \ref{Cosmology} and \ref{DySA}. Sec.\ref{Cosmology} deals with the calculations of cosmologically important parameters, whereas Sec.\ref{DySA} deals with the qualitative analysis of dynamics considered in the former. In Sec.\ref{Cosmology} we start with setting up the equations of motions in LQC framework for universe sourced by a scalar field in an FLRW background. This is followed by introducing few cosmologically important parameters to be used for our analysis. Introducing the form of potential in Sec.\ref{Cosmology}, we further divide the section into two subsections \ref{LinearMonodromy} and \ref{QuadraticMonodromy} for two different potentials: one dominated by linear and the other with quadratic monodromy term respectively as mentioned above. Wherein we calculate the number of e-folds generated during the slowly rolling phase of the universe for both the potentials and plot the relevant parameters for each cases. 
In addition to this we calculate the effects on the number of e-folds produced due to the various free parameters pertaining to the monodromy potential considered in this work. In Sec.\ref{DySA} we carry out a dynamical system analysis. This is broadly divided into two major subsections \ref{DySA1} and \ref{DySA2} for linear and quadratic monodromy potential respectively. These subsections consist of producing the phase portraits and carrying out the stability analysis of the fixed points for each of the cases. In Sec.\ref{DR} we discuss our results and conclude.
 
\section{Dynamics of the Background Cosmology}\label{Cosmology}

 In this section we present and study the evolution of universe in the framework of LQC in a flat FLRW background with a $R^3$ topology. The quantum corrected Friedmann equation is given by:
 
\begin{equation}
H^2= \frac{8 \pi G}{3} \rho \left(1 - \frac{\rho}{\rho_c}\right),
\label{FD}
\end{equation} 
where $H\equiv \frac{\dot{a}}{a}$ and $\rho$ are the Hubble parameter and  the energy density respectively. Whereas, $G=\frac{1}{m^2_{Pl}}$, is the Newton Gravitational constant and $\rho_c$ is the critical energy density. The fact that the correction term $-\frac{\rho^2}{\rho_c}$ is purely due to quantum geometric effects can be seen from the definition of $\rho_{c}=\frac{18\pi}{G^2 \hbar \bigtriangleup^3}$, where $\hbar$ is Planck constant and $\bigtriangleup$ is the area gap \cite{EigenValues}. It is clear that for large eigenvalues and hence in the classical limit $\rho_c$ tends to infinity and we recover the GR case. The important point is that the correction term comes with a negative sign which allows for the occurrence of bounce without the violation of any energy condition unlike that of GR. The conservation of the energy momentum tensor gives the following continuity equation
 
\begin{equation}
\frac{d \rho}{d t} + 3H (\rho + P) = 0,
\label{CEqn}
\end{equation}
where $P$ is the pressure \footnote{In this paper we ignore the inverse-volume corrections in the effective Hamiltonian as the spacetime can be treated classically for large eigenvalues, which otherwise would lead to a different equation of state, and hence a modified continuity equation \cite{InverseVolume}.}. This is exactly the same equation as that of GR. Thus the quantum geometry effects only influence the dynamics of the scalar field through the Hubble parameter $H.$ Now for a universe sourced by a minimally coupled single scalar field with a standard kinetic term and potential $V(\phi)$, as in the case considered here, the continuity equation (\ref{CEqn}) becomes the Klien Gordon Equation:
 
\begin{equation}
\ddot{\phi}+3H\dot{\phi}+V'(\phi)=0,
\label{KG}
\end{equation}
where we have used that $\rho= \frac{1}{2}\dot{\phi^2}+V(\phi)$ and $P= \frac{1}{2}\dot{\phi^2}-V(\phi)$. Whereas dot represents derivative w.r.t the cosmic time, on the other hand, prime represents derivative w.r.t. $\phi$. From Eqs.(\ref{CEqn}) and  (\ref{FD}) it is straight forward to obtain the quantum corrected Raychaudhuri equation using the identity  $\frac{dH}{dt}= \frac{\ddot{a}}{a}-\left(\frac{\dot{a}}{a}\right)^2$ to give: 

\begin{equation}
\frac{\ddot{a}}{a}= -\frac{4 \pi G}{3} \rho \left(1-4 \frac{\rho}{\rho_c}\right)-4 \pi G P \left(1- 2 \frac{\rho}{\rho_c}\right).
\label{RCE}
\end{equation}

A bouncing scenario is obtained as when an initially contracting universe goes to an expanding one through a minimum in the scale factor but not zero. Since the Hubble parameter is proportional to fractional rate of change of volume $H \propto \frac{ \triangle V}{V \triangle t}$, therefore, a contracting (expanding) phase is specified by a negative (positive) value of the Hubble parameter $H.$ Therefore $H$ must pass from negative to positive through zero at the bounce. However, this is a necessary but not sufficient condition. Along with this, the slope of the slope of the scale factor must be positive at the turn around point to achieve the minima and hence the bounce. Mathematically, the following two conditions must be satisfied for the bounce to occur: 

\begin{eqnarray}
\left(\frac{da}{dt}\right)_B &=& 0 , \label{BC1}\\ \newline 
\left(\frac{d^2 a}{dt^2}\right)_B &>& 0. \label{BC2}
\label{BouncingConditions}
\end{eqnarray}

From here onwards the subscript $B$ will be used to denote the point of occurrence of the bounce. It is clear that the first condition of bounce Eq.(\ref{BC1}) is satisfied, in the case of LQC, when the energy density of the matter field reaches the critical value $\rho_c$ as is obvious from Eq.(\ref{FD}). Now substituting the energy density $\rho$ and pressure $P$ into the Raychaudhuri equation (\ref{RCE}) and evaluating it at $\rho=\rho_c$, it is straight forward to see that $\frac{\ddot{a}}{a}= 4 \pi G {\dot{\phi}^2}$ which is always positive definite. And hence the quantum bounce is guaranteed at $\rho = \rho_c.$ We will be synonymously using the word bounce for quantum bounce in this paper. 

Now, to study the dynamics of the universe it is sufficient to consider the evolution of the equations (\ref{FD}) and (\ref{KG}). It is obvious from above that the modified Raychaudhuri equation  (\ref{RCE}) can be obtained from equations (\ref{FD}) and (\ref{CEqn})  and hence it is redundant. As far as the initialization of the system of equations (\ref{FD}) and (\ref{KG}) is concerned, the space of initial data is four dimensional. This includes the initial values of the scale factor $a$, the Hubble parameter $H$, $\phi$ and $\dot{\phi}.$ However the value of the scale factor at initial time enjoys  a constant rescaling freedom without altering the physical results. Utilizing this freedom one can fix the scale factor at the bounce $a_B=1$ without loss of generality. The choice of the bounce point as the initial condition further provides the initial value of the Hubble parameter $H=0$ satisfying the first condition Eq.(\ref{BC1}) at the  bounce where $\rho=\rho_c$ \footnote{In this paper, we set up the initial conditions at the moment of the bounce \cite{AS11}. In the literature, other choices of the initial moment exist, and one of them is  the remote past of the bounce, see in particular the interesting arguments of \cite{MBS17}.}. This, $\rho_B= \frac{1}{2}\dot{\phi^2_B}+ V(\phi_B)  =\rho_c$, in turn determines the initial value of $\dot{\phi}$ up to a sign once the initial value of $\phi$ is fixed. Hence, with $a_B$ fixed the value of $\phi_B$ and the sign of $\dot{\phi_B}$ completely determine the space of the initial data. In fact the set of initial conditions is the locus of all points on the phase space of $\phi$ and $\dot{\phi}$ satisfying the condition $\rho_c=\frac{1}{2}\dot{\phi^2}+V(\phi).$ In this work we concentrate on an initially kinetic dominated universe with positive initial $\dot{\phi}$ for all the cases to be discussed below.

\begin{figure}
$
 \begin{array}{c c}
   \includegraphics[width=0.44\textwidth]{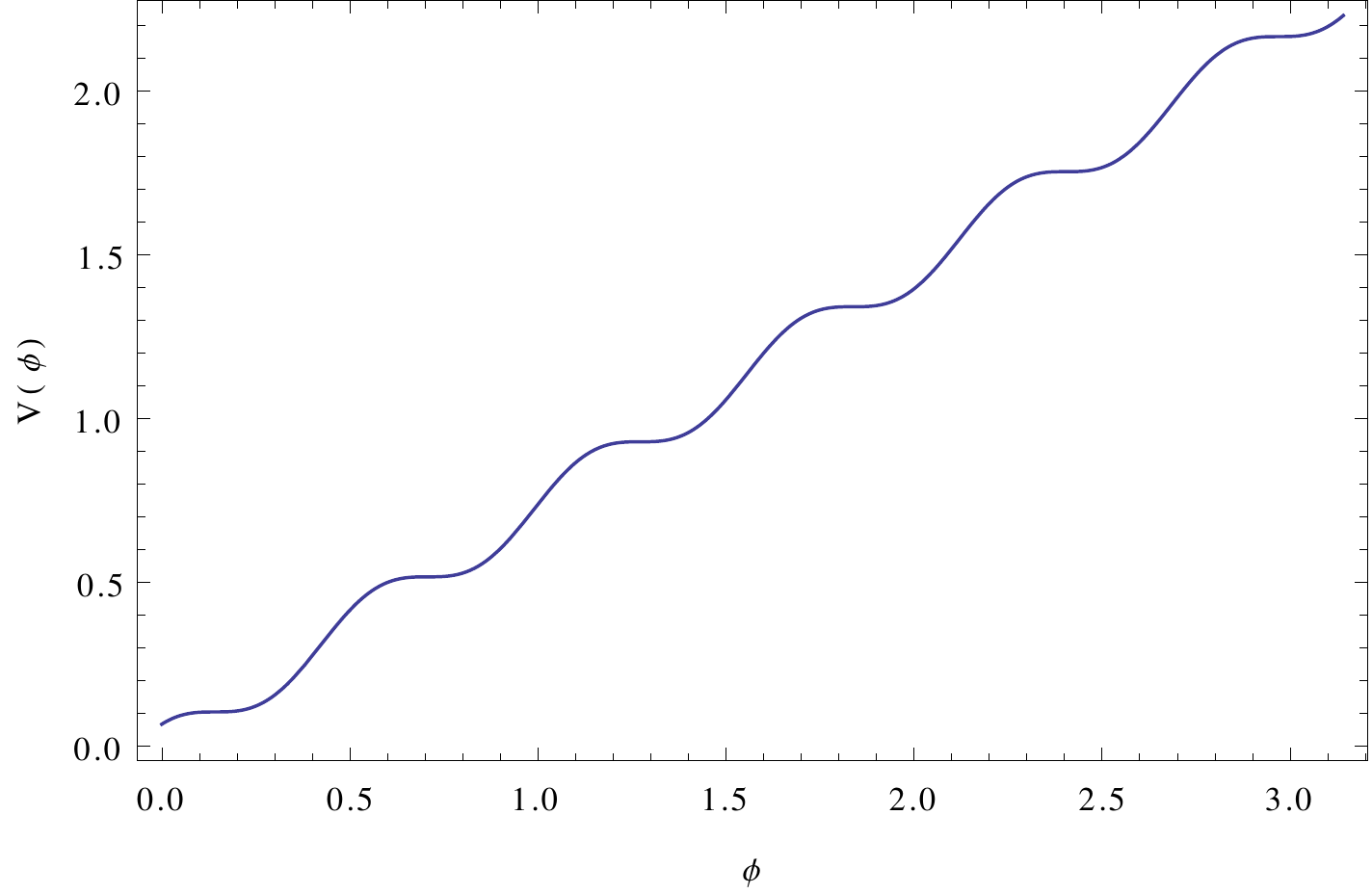} &
   \includegraphics[width=0.436\textwidth]{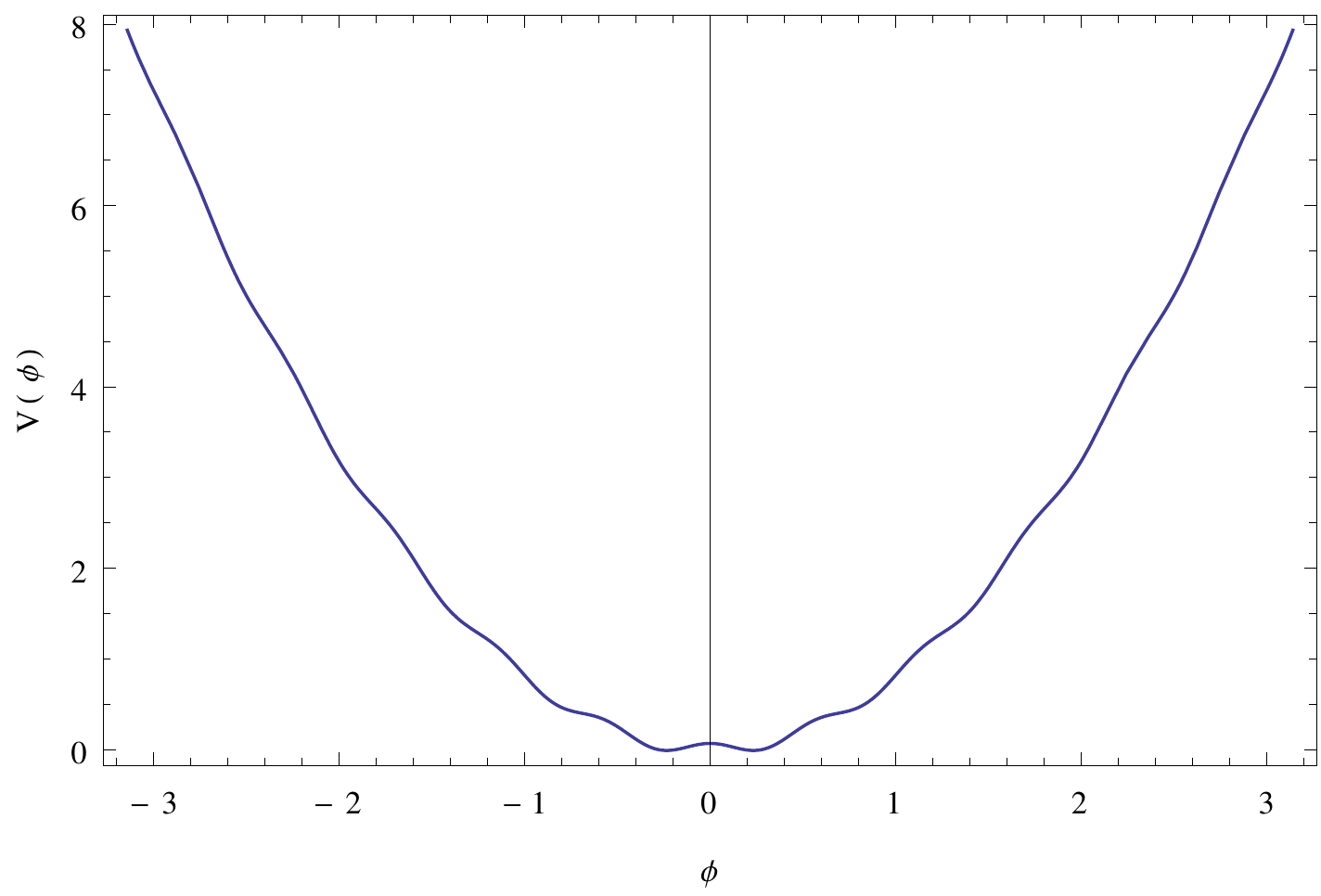} \\
   
 \end{array}
 $
  
  \caption{The \emph{left plot} is for linear monodromy case of the form $V(\phi)= \mu^3 [\phi + bf cos(\frac{\phi}{f} +\delta)]$   and the \emph{right plot} for quadratic monodromy case of the form $V(\phi)= \mu^3 [\frac{\phi^2}{\mu} + bf cos(\frac{\phi}{f} +\delta)]$ with the set of parameters $(b,f,\delta)=(0.99,0.09,0)$ and $\mu=0.9$ for both the plots.}
 \label{Potentials} 
\end{figure}

Now the potential with a monomial monodromy and modulation takes the form \cite{MonodromyPotential1}:
\begin{equation}
V(\phi)= {\mu}^{4-p} {\phi}^p + \sum_i \Lambda_i^4 cos\left(\frac{\phi}{f_i}+\delta_i \right), \label{Potential}
\end{equation}
where the first term in Eq.(\ref{Potential}) is the monodromy potential and second being a modulation. The $\mu$, $\Lambda$, $f_i$ and  $\delta_i$ are, respectively, a constant mass factor, strength of modulation, decay constant and a constant phase factor whereas $p$ is a real number. In this paper we will study background dynamics of an FRW universe sourced by a  minimally coupled scalar field with monodromy potential with only one modulation term in loop quantum cosmology framework with two different values of $p=1$ and $2.$  Therefore our potential looks like: 
\begin{equation}
V(\phi)= {\mu}^{4-p} {\phi}^p +  \Lambda^4 cos\left(\frac{\phi}{f}+\delta\right). \label{Potential}
\end{equation}
In the following we express the potential in an elegant way useful for our calculation.
\begin{eqnarray} \nonumber
V(\phi) &=& {\mu}^{4-p} {\phi}^p + \Lambda^4 cos\left(\frac{\phi}{f}+\delta\right) \\ 
        &=& {\mu}^3 \left[ \frac{{\phi}^p}{\mu^{p-1}} + bf cos\left(\frac{\phi}{f}+\delta\right) \right],
\end{eqnarray}
where $b= \frac{\Lambda^4}{{\mu}^3 f}$. The values of parameters $b,f$ must be less than 1 and while $\delta$ can take values in the range $[0, 2\pi]$ \cite{MonodromyPotential1}. The profile of the potential for $p=1$ and $2$ are shown in Fig.\ref{Potentials}. In this paper when  we will say linear (quadratic) case, it is always to be understood as a linear (quadratic) monodromy potential with a small modulation term. Also note that $\mu$, $\Lambda$ and $f$ having the dimension of mass makes $b$ a dimensionless parameter. In this work we set $m_{Pl}=1$ for all our numerical simulations.

The background evolution has been extensively investigated and one of the robust result is that following the bounce a desired slow roll inflation is most likely, provided that the kinetic energy of the scalar field is dominant over the potential energy initially at the quantum bounce. To this effect, we introduce the following parameters: 

(1) The equation of state (EOS), $w$: It is defined as
\begin{equation}
w(\phi)= \frac{\frac{1}{2} \dot{\phi^2} - V(\phi)}{\frac{1}{2}\dot{\phi^2}+V(\phi)}.
\end{equation}
It can be immediately inferred from the above that for a slowly rolling inflaton field, $V(\phi)>>\frac{1}{2}\dot{\phi^2},$ and hence $w(\phi) \approx -1$ to give an almost exponential expansion.

(2) The inflation is characterized by acceleration of the universe $\ddot{a}>0$. However it is important to have a prolonged period of inflation to produce enough number of e-folds to cure the problems of the big bang cosmology. This is often designated as slowly rolling phase of inflation when the potential energy dominates over the kinetic energy, $V(\phi)>>\frac{1}{2}\dot{\phi^2}$, and the inflaton rolls slowly enough such that $\ddot{\phi}\simeq 0.$ These conditions can be parametrized in terms of $e_V$ and $\eta_V$ called slow roll parameters defined as 

\begin{eqnarray}
\epsilon_V = \frac{1}{16 \pi G} \left(\frac{V'(\phi)}{V(\phi)}\right)^2, \hspace{0.35cm}
\eta_V = \frac{1}{8 \pi G}   \frac{V''}{V}.
\label{SRP}
\end{eqnarray}
The slow roll inflation is attained as long as $\epsilon_V,$ $|\eta_V|<<1.$

(3) The number of e-folds, $N_{inf}$: The amount of expansion of the universe during inflation is quantified as $N_{inf}= ln\left(\frac{a_{f}}{a_i}\right)= \int^{t_f}_{t_i} H dt \simeq \int^{\phi_f}_{\phi_f}\frac{H}{\dot{\phi}}d\phi.$ While slow roll conditions are satisfied, this becomes

\begin{equation}
N_{inf}= \int^{\phi_i}_{\phi_{end}}\frac{V}{V'(\phi)}d{\phi}.
\end{equation}
In this article we will use the subscript ``i" and ``f" to denote, respectively, the starting and end time of slow roll inflation. As per convention in \cite{o}, we use $\ddot{a}(t=t_i)=0$
and $w(t=t_f)=-\frac{1}{3}$ to obtain $t_i$ and $t_f$ (See also \cite{SSW}). Though, strictly speaking, these choice of time is not in accordance with the definition of slow roll inflation but it is safe to use them because it is insensitive to observations\cite{UnivFeatA,UnivFeatB}.

(4) An useful parameter to quantify the dominance of either kinetic energy or potential energy is via $r_w= \frac{K.E.}{P.E.}=\frac{\dot{\phi^2}/2}{V(\phi)}$. It is obvious that for the case of potential energy dominance (PED) $r_w<1$ and $r_w>1$ for kinetic energy dominance case (KED).  
In particular we are interested  to obtain the critical value of $r_w^c$ that would generate $60$ e-folds of expansion of the universe in the slow roll regime. The correspondence between $w_{\phi}$ and $r_w$ is shown below:

\begin{equation}
w_{\phi} = \frac{r_w -1}{r_w +1}.
\label{Relation} 
\end{equation}
It is clear from above that for the case of KED ($r_w>1$) we have $w_{\phi}>0$ while for PED ($r_w<1$) we have $w_{\phi}<0.$

\subsection{Background with Linear Monodromy Potential: $V(\phi)= \mu^3 \left[\phi + bf cos\left(\frac{\phi}{f} +\delta\right)\right]$}\label{LinearMonodromy}

Here we present and analyze our results of a Monodromy potential of the linear form with a modulation in LQC setup. We show the evolution of the cosmologically important parameters discussed in the previous section \ref{Cosmology} and calculate  the number of e-folds $N_{inf}$ to this effect.
\begin{figure}
$
 \begin{array}{c c}
   \includegraphics[width=0.44\textwidth]{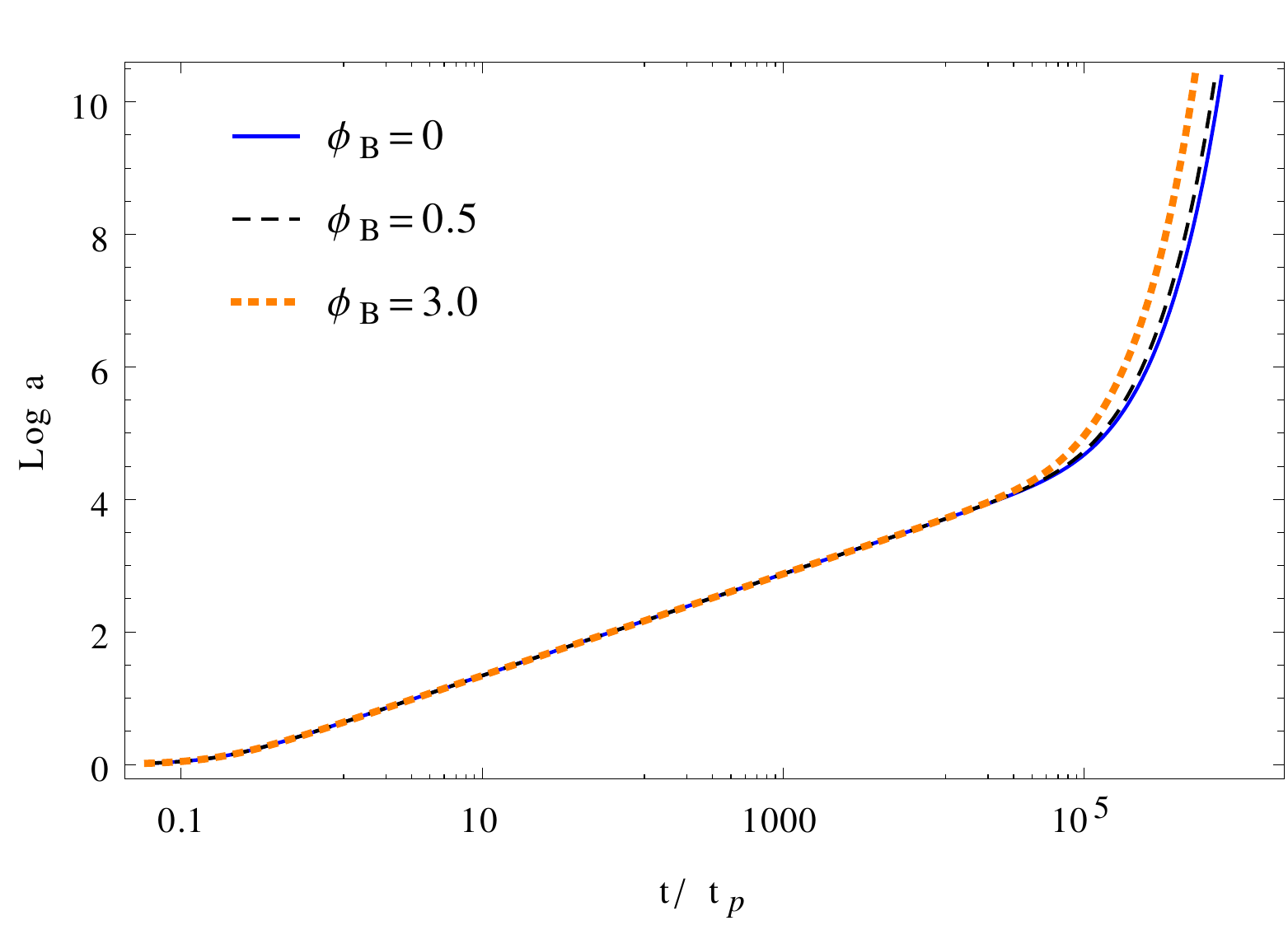} &
   \includegraphics[width=0.463\textwidth]{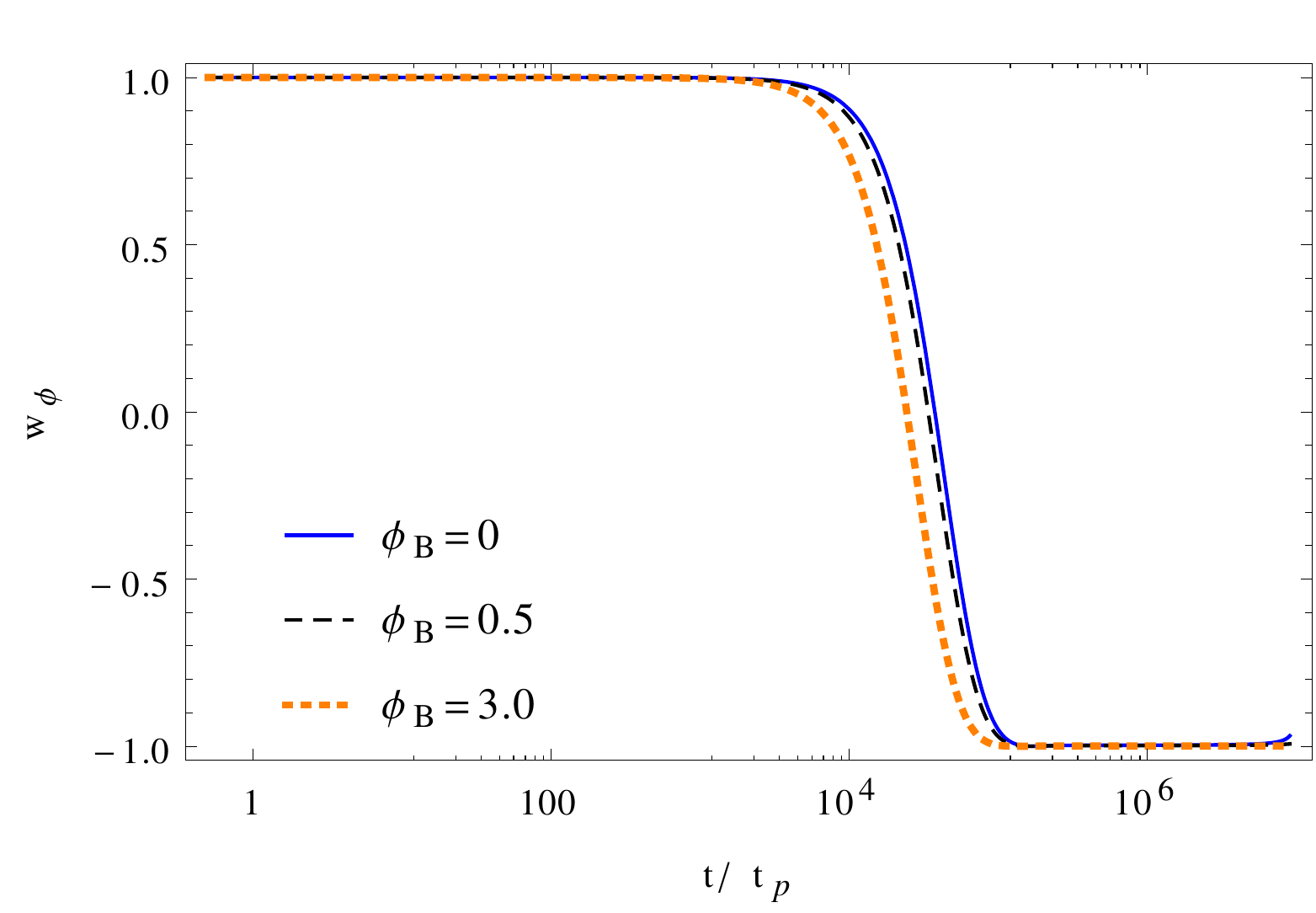} \\
   
 \end{array}
 $
  
  \caption{The \emph{left plot} is for $Log$ $a(t)$ and the \emph{right plot} for $w_{\phi}$ with $b=0.001$, $f=0.1$ and $\delta=0$ for linear monodromy potential of the form $V(\phi)= \mu^3 [\phi + bf cos(\frac{\phi}{f} +\delta)]$ with $\mu =  1.51152 \times 10^{-4}m_{Pl}$ and setting $m_{Pl}=1$.}
 \label{Fig1} 
\end{figure}

\begin{figure}
$
 \begin{array}{c c}
   \includegraphics[width=0.45\textwidth]{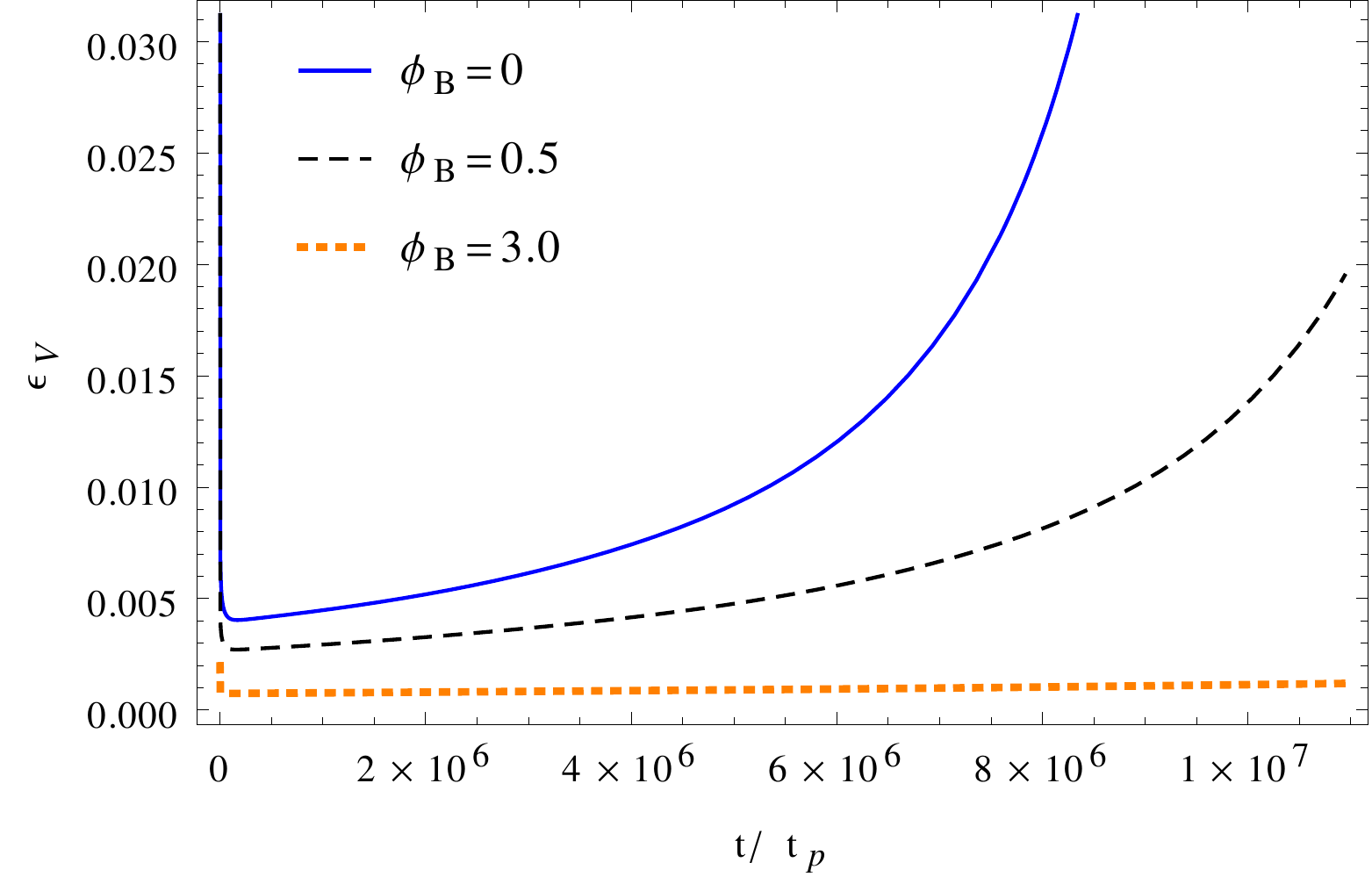} &
   \includegraphics[width=0.472\textwidth]{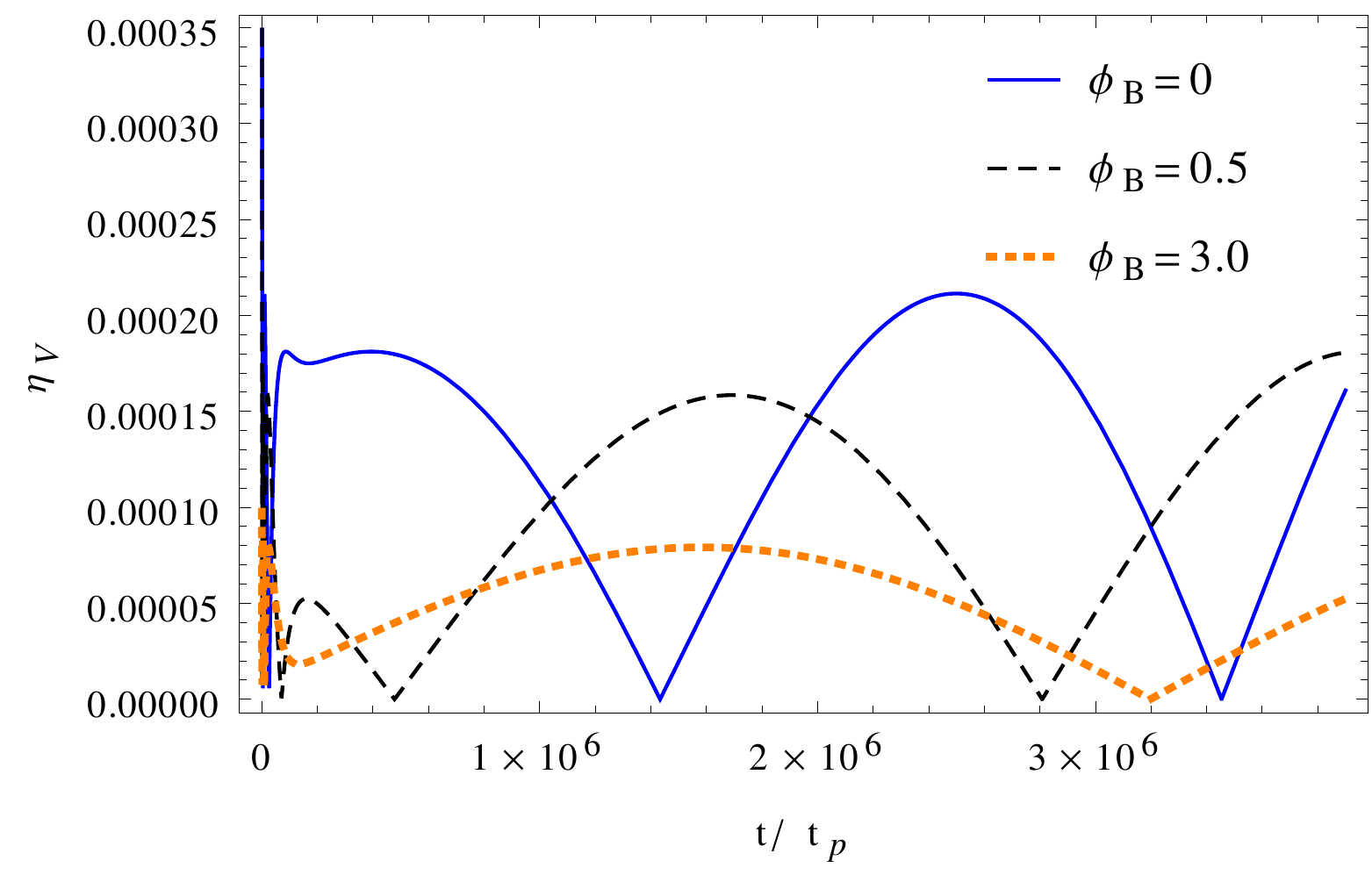} \\
   
 \end{array}
 $
  
  \caption{The \emph{left plot} is for $\epsilon_V$ and the \emph{right plot} for absolute value of $\eta_V$ with $b=0.001$, $f=0.1$ and $\delta=0$ for linear monodromy potential of the form $V(\phi)= \mu^3 [\phi + bf cos(\frac{\phi}{f} +\delta)]$ with $\mu = 1.51152 \times 10^{-4} m_{Pl} $ and setting $m_{Pl}=1$.}
 \label{Fig2} 
\end{figure}

We use the Friedmann Eq.(\ref{FD}) and the Klein Gordon Eq.(\ref{KG}) to numerically simulate the background dynamics. All the figures produced are for initially kinetically dominated universe and w.r.t cosmic time expressed in units of Planck time. In figures \ref{Fig1} and \ref{Fig2} we show the relevant dynamical quantities of the background with linear monodromy potential of the form $V(\phi)= \mu^3 [\phi + bf cos(\frac{\phi}{f} +\delta)]$ with values of parameters $(b,f,\delta)=(0.001,0.1,0)$ for different initial conditions of $\phi_B.$ We initialize a kinetically dominated universe at the quantum bounce. In the left of Fig.\ref{Fig1} we plot the natural logarithmic of the scale factor $a(t)$ and on the right the equation of state parameter $w_{\phi}$. The explicit expression of $w_{\phi}$ for a linear monodromy potential with a modulation term looks like
\begin{equation}
w(\phi) = \frac{\frac{\dot{\phi}^2}{2}-\mu^3\left[ \phi + bf cos\left( \frac{\phi}{f} +\delta\right)\right]}{\frac{\dot{\phi}^2}{2} + \mu^3\left[ \phi + bf cos\left( \frac{\phi}{f} +\delta\right)\right]}.
\label{wphiLinear}\end{equation}
Coming to the analysis of Fig.\ref{Fig1}, the parameter $w(\phi)$ starts with almost $1$ and then slowly transits to $-1.$ This can be clearly seen from the expression above Eq.(\ref{wphiLinear}). For small values of $\phi_B$ the potential term in both the numerator and denominator of Eq.(\ref{wphiLinear}) can be neglected in comparison to the kinetic term calculated using Eq.(\ref{BC1}). To be specific the smallness of the value of the potential term is justified as the values of the product $bf$ is always set to less than unity. Thus the equation of state parameter $w_{\phi}\simeq 1$ initially. This behavior of the energy-momentum  tensor is also called stiff fluid in the literature. As the universe evolves it transits from $w_{\phi}\simeq 1$ to $\simeq -1$ which represents the slowly rolling phase of the inflation to be seen below. This can be easily understood from Eq.(\ref{wphiLinear}). As the potential energy dominates over the kinetic energy $w_{\phi}$ becomes $-1.$ No matter with what value of initial values of $\phi_B$ we start out with, the universe gets attracted to the $w_{\phi}=-1$ solution for the KED cases. Although the different values of initial conditions $\phi_B$ will affect the duration of the inflationary phase to be discussed below. As have been vastly reported in the literature, it is also true for our choice of potential that there exist three phases of evolution, namely, bouncing, transition and inflation. The behavior of the log of scale factor $a(t)$ depicts the universality of the solutions in the bouncing phase for kinetic energy dominant initial conditions at the quantum bounce. 

Now coming to the analysis of Fig.\ref{Fig2}, the two slowly rolling parameters, $\epsilon_V$ and the absolute value of $\eta_V$  are plotted. A slow roll inflation is guaranteed if $|\eta_V|$,$\epsilon_V<<1.$ And Fig.\ref{Fig2} shows the satisfaction of these two conditions. For both the parameters in Eq.(\ref{SRP}) it can be easily seen that when the potential energy dominates and varies very slowly w.r.t the scalar field $\phi$, the values of $\epsilon_V,$ $|\eta_V|<<1$ and Fig.\ref{Fig2} exactly portraits it. This is the regime where the slow roll inflation happens untill and unless $\epsilon_V, |\eta_V|=1.$ The right of Fig.\ref{Fig2} for absolute values of $\eta_V$ shows some  oscillations, but its amplitudes are very much less than unity to assure the slow roll phase of the inflation. Thus in Fig.\ref{Fig2} we explicitly show the slow roll regime of the evolution of the universe.

One of the major goals of this paper is to calculate the number of e-folds $N_{inf}.$ In order to parametrize the dominance of energy we introduced a quantity called $r_w$ in section \ref{Cosmology}. Since from the observation it suggests that a $60$ number of e-folds are necessary to solve the puzzles of the standard model, therefore, we are interested to know for  what initial value of $\phi_B$ and hence $r_w$ would the universe land up with this number. As mentioned in Sec.\ref{Introduction} this value of $r_w$ is designated as $r_w^c$ following \cite{o}. To meet this we calculate the number of e-folds using the criteria for start and end of inflation as mentioned in Sec.\ref{Cosmology}. Details of the calculations are tabulated in Table \ref{table1}. Once the values of the parameters $b$ and $f$ are fixed, the only free parameter is the value of $\phi$ at the bounce as $\dot{\phi}$ is determined upto a sign due to Eq.(\ref{BC1}). In this work we focus only on the positive value of the initial velocity of the field. Thus, Table \ref{table1}  lists the details of the calculation including the starting and ending values of time for inflation and the inflaton $\phi$ for various $\phi_B$ fixing the parameters $b=0.001$, $f=0.1$ and $\delta=0$. From Table \ref{table1} it is evident that more than the desired number of e-folds, $N_{inf}\simeq 64,$ is achieved even we start with $\phi_B=0$. Now when we increase $\phi_B$ the number $N_{inf}$ also increases. Therefore, there is no critical $r_w$ here as more than $60$ e-folds is always obtained. 
 
Next, we consider studying the effects of variation of the parameters $b$ and $f$ on the value of $N_{inf}$ in Table \ref{table2}. It has been found in our analysis not shown here that the change in the value of $N_{inf}$ is negligible if we individually vary either the parameter $b$ or $f$ but not both at time. It is because of the fact that it is the product of $b$ and $f$ that appears as a coefficient of the sinusoidal term in the potential. The dynamics is least effected on individual variation of these two parameters as long as they are less than one. Therefore, we show here the effects of variation of the product $bf$ in Table \ref{table2}. It is clear that the number $N_{inf}$ increases with increase in value of the product $bf$ keeping $\phi_B$ fixed. And this exercise is repeated for different values of $\phi_B$ to arrive at the same conclusion. Thus we can infer that the value of $bf$ can be tuned in to get the desired number of $N_{inf}$. In the limit $bf\rightarrow 0$ we will recover the case of power law potential with linear form \cite{o}\cite{ro}.
 
Next let us consider the variation of the parameter $\delta$ on $N_{inf}.$ In order to do this 
we set the values of the parameters $(b,f)=(0.9,0.9)$ to increase the strength of the modulation term. To our observations we find that the value of the phase factor $\delta$ has an adverse effect on the value of $N_{inf}.$ The number of e-folds generated during the slowly rolling phase of the universe is drastically altered if we change the value of $\delta$ in Table \ref{table3}. We attribute this to the properties of the sinusoidal function in the modulation term which can change its value and sign depending upon the value of the argument which here is the phase factor $\delta$.

\begin{table}

\begin{center}

\begin{tabular}{ccccc}
\hline\hline
 $\phi_B$~~~  & Inflation~~~ & $t/t_{pl}$~~~ & $\phi_{*}$~~~ & $N_{inf}$ \\
\hline\hline
\\
 0~~~&starts~~~& $4.811 \times 10^4$ ~~~& 2.142~~~ & 63.99 \\
 ~~~&ends~~~& $1.063 \times 10^7$ ~~~& 0.07764~~~ &  \\\\
 
\\
 0.001~~~&starts~~~& $4.81 \times 10^4$ ~~~& 2.127~~~ & 64.499 \\
 ~~~&ends~~~& $1.064 \times 10^7$ ~~~& 0.07636~~~ &  \\\\
 
 \\
 0.01~~~&starts~~~& $4.8001 \times 10^4$ ~~~& 2.135~~~ & 65.194 \\
 ~~~&ends~~~& $1.069 \times 10^7$ ~~~& 0.08141~~~ &  \\\\
 
 \\
 0.05~~~&starts~~~& $4.758 \times 10^4$ ~~~& 2.191~~~ & 66.943 \\
 ~~~&ends~~~& $1.09 \times 10^7$ ~~~& 0.0773~~~ &  \\\\
 
 \\
 0.1~~~&starts~~~& $4.707 \times 10^4$ ~~~& 2.24~~~ & 70.366 \\
 ~~~&ends~~~& $1.131 \times 10^7$ ~~~& 0.07972~~~ &  \\\\
 
 \\
 0.5~~~&starts~~~& $4.352 \times 10^4$ ~~~& 2.628~~~ & 93.765 \\
 ~~~&ends~~~& $1.418 \times 10^7$ ~~~& 0.00.0806~~~ &  \\\\
 
 \\
 1.0~~~&starts~~~& $4.003 \times 10^4$ ~~~& 3.093~~~ & 132.69 \\
 ~~~&ends~~~& $1.809 \times 10^7$ ~~~& 0.08047~~~ &  \\\\
 
 \\
 2.0~~~&starts~~~& $3.496 \times 10^4$ ~~~& 4.069~~~ & 219.787 \\
 ~~~&ends~~~& $2.678 \times 10^7$ ~~~& 0.0805~~~ &  \\\\
 
 \\
 3.0~~~&starts~~~& $3.142 \times 10^4$ ~~~& 2.127~~~ & 336.799 \\
 ~~~&ends~~~& $3.674 \times 10^7$ ~~~& 0.07636~~~ &  \\\\
 
\hline\hline
\end{tabular}

\end{center}
\caption{This table shows the values of the number of e-folds, $N_{inf}$, for various initial values $\phi_B$ with fixed $b=0.001$ and $f=0.1$ for potential of the form $V(\phi)= \mu^3\left[ \phi + bf cos\left(\frac{\phi}{f}+ \delta\right)\right]$ with $\delta=0$ and $\mu=1.51152 \times 10^{-4}m_{Pl}$. }\label{table1}
\end{table}

\begin{table}

\begin{center}

\begin{tabular}{cccc}
\hline\hline
 $\phi_B$~~~  & $(b,f)=(0.001,0.1)$~~~ & $(b,f)=(0.095,0.79)$~~~ & $(b,f)=(0.9,0.9)$ \\
\hline\hline
\\
$0$ ~~~&$N_{inf}=63.99$~~~& $N_{inf}=70.1$ ~~~& $N_{inf}=364.14$ \\
 ~~~&$t_i= 4.811 \times 10^4$~~~& $t_i= 4.8869 \times 10^4$ ~~~& $t_i= 5.6522 \times 10^4$ \\
  ~~~&$t_f= 1.063 \times 10^7$~~~& $t_f= 1.164 \times 10^7$ ~~~& $t_f= 5.827 \times 10^7$
\\\\

\\
$0.1$ ~~~&$N_{inf}=69.33$~~~& $N_{inf}=75.672$ ~~~& $N_{inf}=375.219$ \\
 ~~~&$t_i= 4.811 \times 10^4$~~~& $t_i= 4.7819 \times 10^4$ ~~~& $t_i= 4.0012 \times 10^4$ \\
  ~~~&$t_f= 1.063 \times 10^7$~~~& $t_f= 1.123 \times 10^7$ ~~~& $t_f= 5.952 \times 10^7$
\\\\ 

\\
$2.0$ ~~~&$N_{inf}=217.576$~~~& $N_{inf}=222.374$ ~~~& $N_{inf}=468.173$ \\
 ~~~&$t_i= 3.4807 \times 10^4$~~~& $t_i= 3.4960 \times 10^4$ ~~~& $t_i= 3.5384 \times 10^4$ \\
  ~~~&$t_f= 2.696 \times 10^7$~~~& $t_f= 2.68 \times 10^7$ ~~~& $t_f= 7.033 \times 10^7$
\\\\ 
\hline\hline
\end{tabular}

\end{center}
\caption{This table shows the values of the number of e-folds, $N_{inf}$, w.r.t. the variation of the product of the parameters $bf$ for three different initial values $\phi_B$ for potential of the form $V(\phi)= \mu^3\left[ \phi + bf cos\left(\frac{\phi}{f}+ \delta\right)\right]$ with $\delta=0$ and $\mu=1.51152 \times 10^{-4}m_{Pl}$.}\label{table2}
\end{table}

\begin{table}

\begin{center}

\begin{tabular}{ccccc}
\hline\hline
 $(\phi_B,\delta)$~~~  & $\phi_B=0$~~~ & $\phi_B=0.05$~~~ & $\phi_B=1.0$~~~ & $\phi_B=2.0$ \\
\hline\hline
\\
$\delta=0$ ~~~&$N_{inf}=367.4$~~~& $N_{inf}=371.2$ ~~~& $N_{inf}=422.3$~~~ &$471.9$ \\
 ~~~&$t_i= 5.6523 \times 10^4$~~~& $t_i= 5.6142 \times 10^4$ ~~~& $t_i= 4.5685 \times 10^4$~~~ &$t_i= 3.5375\times 10^4$ \\
  ~~~&$t_f= 5.827 \times 10^7$~~~& $t_f= 5.901 \times 10^7$ ~~~& $t_f= 6.575 \times 10^7$~~~& $t_f= 7.05\times 10^7$
  \\\\
 
 \\
$\delta=\frac{\pi}{4}$ ~~~&$N_{inf}=202.29$~~~& $N_{inf}=205.01$ ~~~& $N_{inf}=233.6$~~~ &$286.04$ \\
 ~~~&$t_i= 5.9875 \times 10^4$~~~& $t_i= 5.8788 \times 10^4$ ~~~& $t_i= 4.2225 \times 10^4$~~~ &$t_i= 3.3096\times 10^4$ \\
  ~~~&$t_f= 4.434 \times 10^7$~~~& $t_f= 4.427 \times 10^7$ ~~~& $t_f= 4.78 \times 10^7$~~~& $t_f= 5.335\times 10^7$
  \\\\
  
  \\
$\delta=\frac{2\pi}{4}$ ~~~&$N_{inf}=33.56$~~~& $N_{inf}=31.26$ ~~~& $N_{inf}=64.09$~~~ &$137.1$ \\
 ~~~&$t_i= 5.4807 \times 10^4$~~~& $t_i= 5.3563 \times 10^4$ ~~~& $t_i= 3.8385 \times 10^4$~~~ &$t_i= 3.2025\times 10^4$ \\
  ~~~&$t_f= 9.409 \times 10^6$~~~& $t_f= 9.632 \times 10^6$ ~~~& $t_f= 1.337 \times 10^7$~~~& $t_f= 1.98\times 10^7$
  \\\\
  
  \\
$\delta=\frac{3\pi}{4}$ ~~~&$N_{inf}=22.3$~~~& $N_{inf}=24.15$ ~~~& $N_{inf}=69.79$~~~ &$215.8$ \\
 ~~~&$t_i= 4.7552 \times 10^4$~~~& $t_i= 4.6640 \times 10^4$ ~~~& $t_i= 3.6125 \times 10^4$~~~ &$t_i= 3.2540\times 10^4$ \\
  ~~~&$t_f= 3.915 \times 10^6$~~~& $t_f= 4.106 \times 10^6$ ~~~& $t_f= 8.752 \times 10^6$~~~& $t_f= 2.148\times 10^7$
  \\\\
  
  \\
$\delta=\frac{4\pi}{4}$ ~~~&$N_{inf}=33.03$~~~& $N_{inf}=35.18$ ~~~& $N_{inf}=116$~~~ &$597.05$ \\
 ~~~&$t_i= 4.2745 \times 10^4$~~~& $t_i= 4.2171 \times 10^4$ ~~~& $t_i= 3.1625 \times 10^4$~~~ &$t_i= 3.4520\times 10^4$ \\
  ~~~&$t_f= 4.828 \times 10^6$~~~& $t_f=  5.073\times 10^6$ ~~~& $t_f= 1.304 \times 10^7$~~~& $t_f= 5.641\times 10^7$
  \\\\
\hline\hline
\end{tabular}

\end{center}
\caption{This table shows the values of the number of e-folds, $N_{inf}$, for five different values of $\delta=(0\pi/4, 1\pi/4, 2\pi/4, 3\pi/4, 4\pi/4)$ and various initial values $\phi_B$ fixing $b=0.9$ and $f=0.9$ for potential of the form $V(\phi)= \mu^3 \left[\phi + bfcos\left(\frac{\phi}{f}+ \delta\right)\right]$ with $\mu=1.51152 \times 10^{-4}$ $m_{Pl}.$}\label{table3}
\end{table}

 \subsection{Background with Quadratic Monodromy Potential: $V(\phi)= \mu^3 \left[\frac{\phi^2}{\mu} + bf cos\left(\frac{\phi}{f} +\delta\right)\right]$}\label{QuadraticMonodromy}
\begin{figure}
$
 \begin{array}{c c}
   \includegraphics[width=0.44\textwidth]{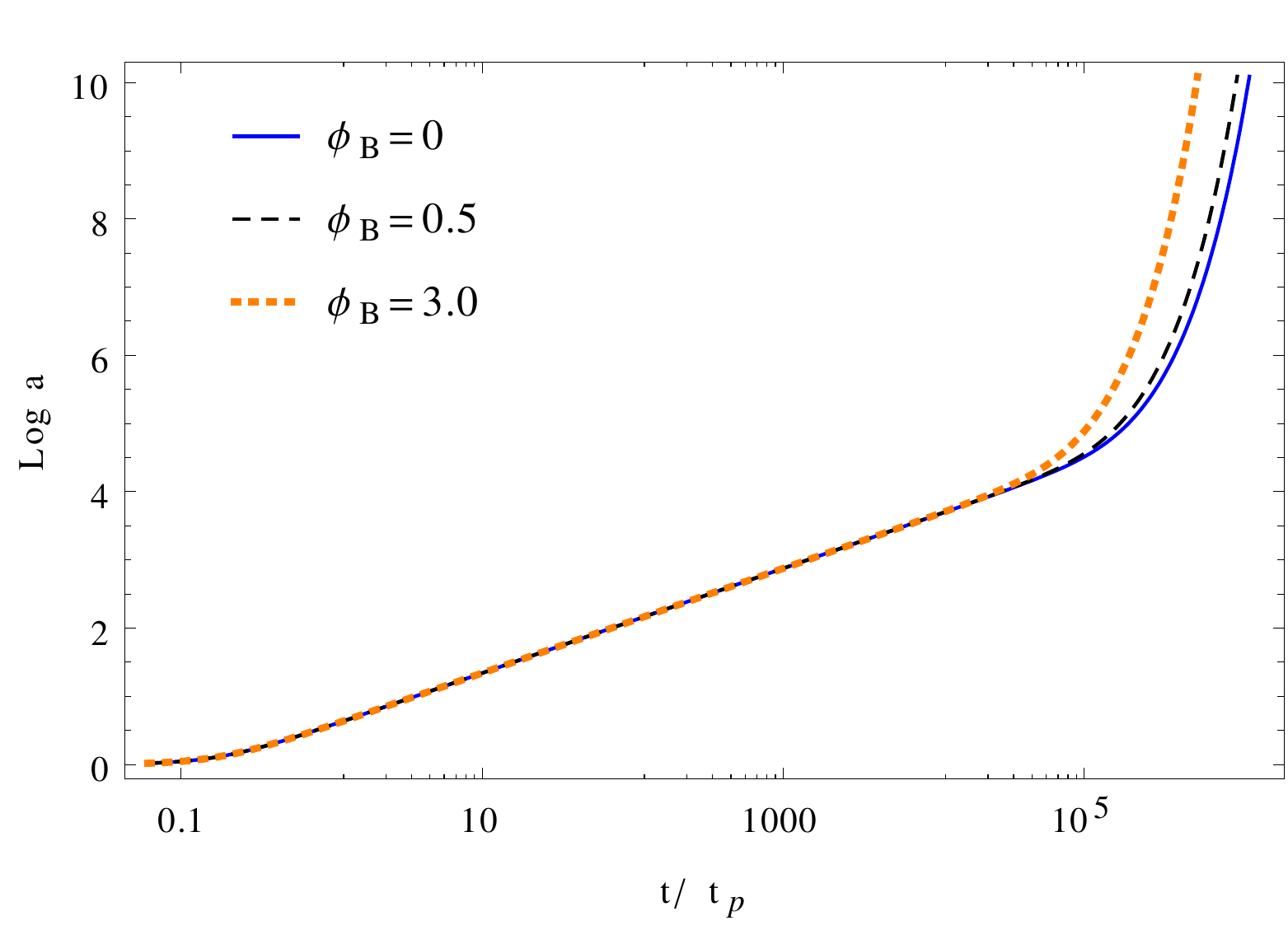} &
   \includegraphics[width=0.463\textwidth]{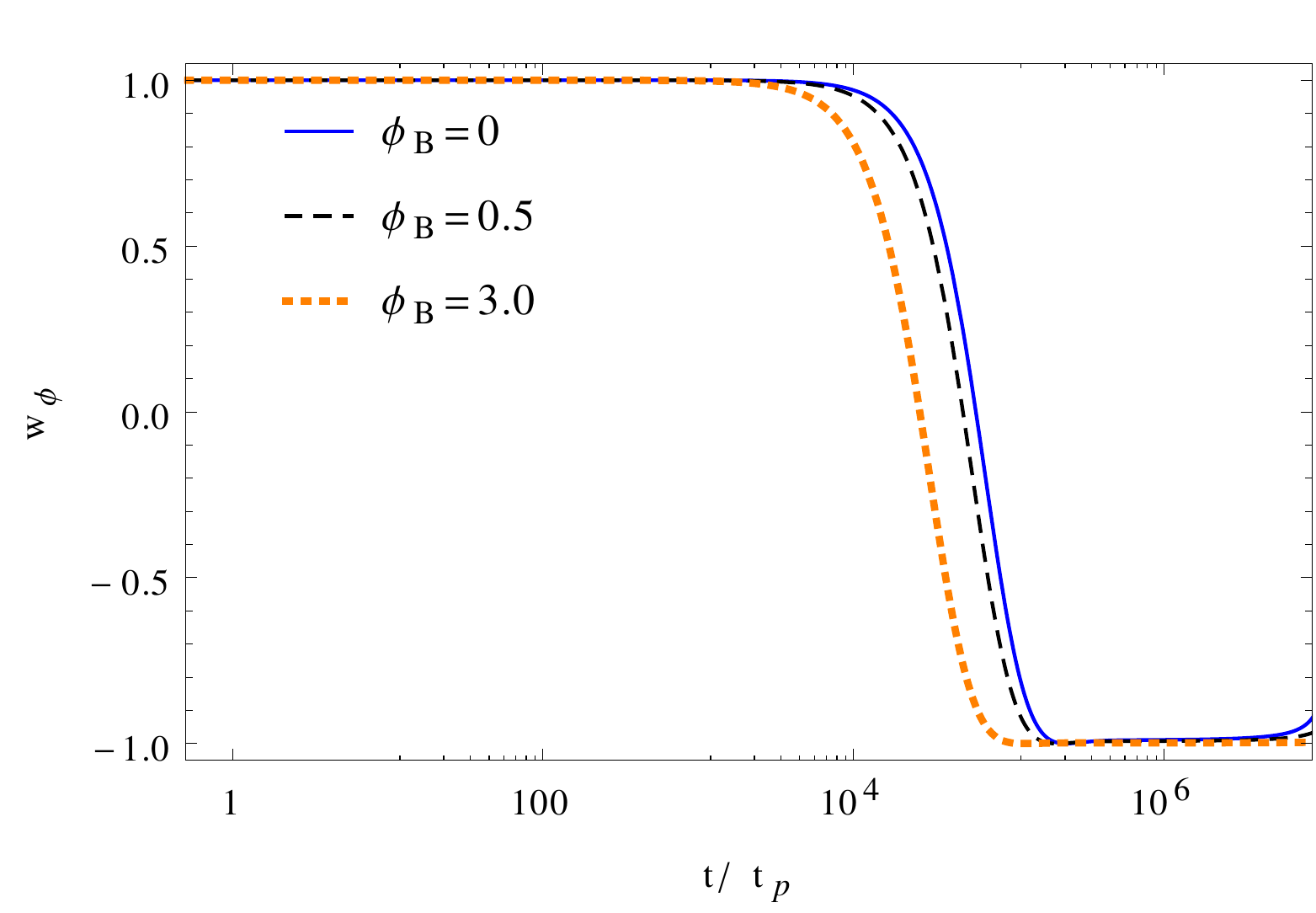} \\
   
 \end{array}
 $
  
  \caption{The \emph{left plot} is for $Log$ $a(t)$ and the \emph{right plot} for $w_{\phi}$ with $b=0.001$, $f=0.1$ and $\delta=0$ for quadratic monodromy potential of the form $V(\phi)= \mu^3 [\frac{\phi^2}{\mu^2} + bf cos(\frac{\phi}{f} +\delta)]$ with $\mu =  0.749533 \times 10^{-6}m_{Pl}$ and setting $m_{Pl}=1$.}
 \label{Fig3} 
\end{figure}

\begin{figure}
$
 \begin{array}{c c}
   \includegraphics[width=0.45\textwidth]{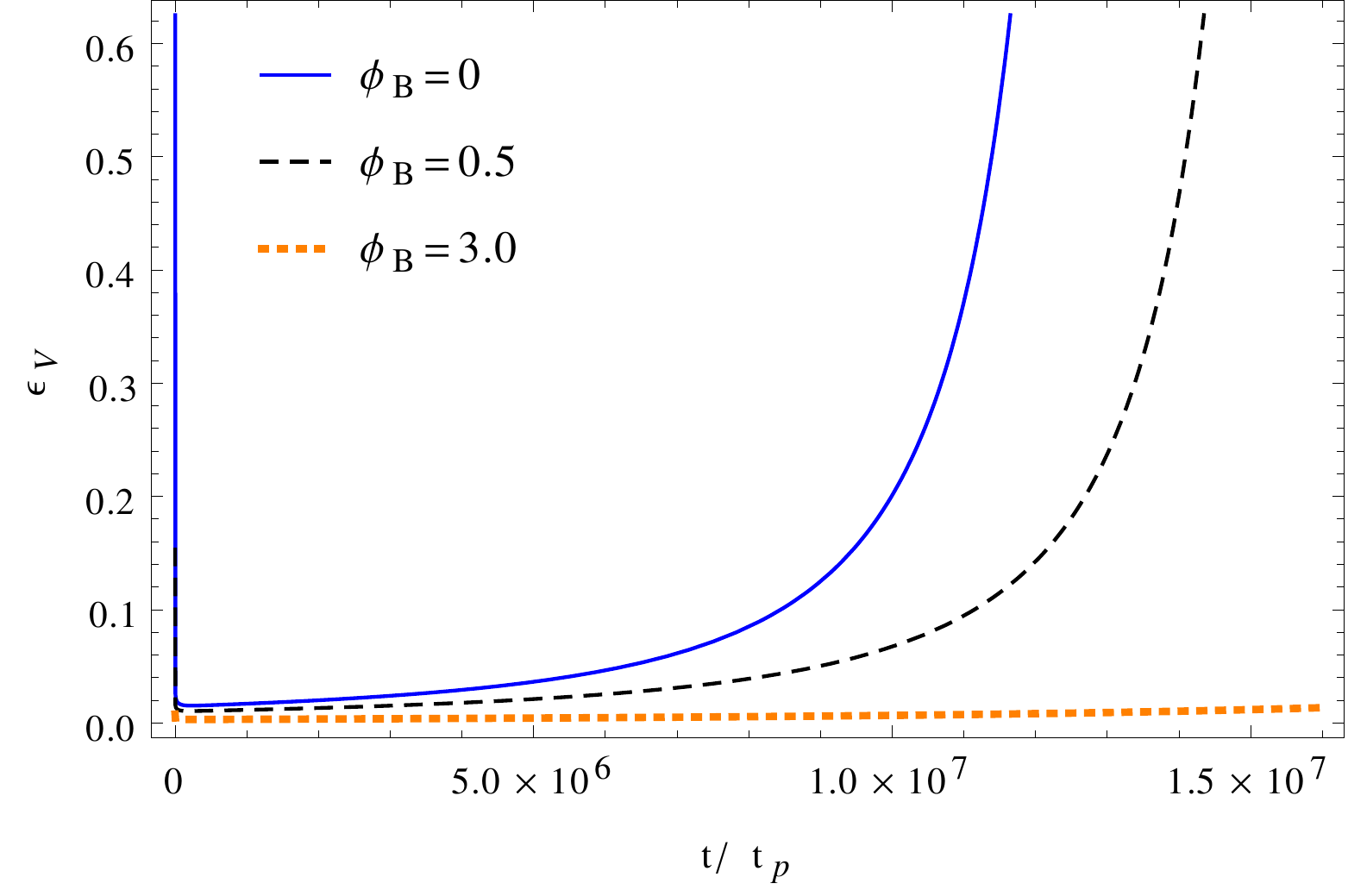} &
   \includegraphics[width=0.45\textwidth]{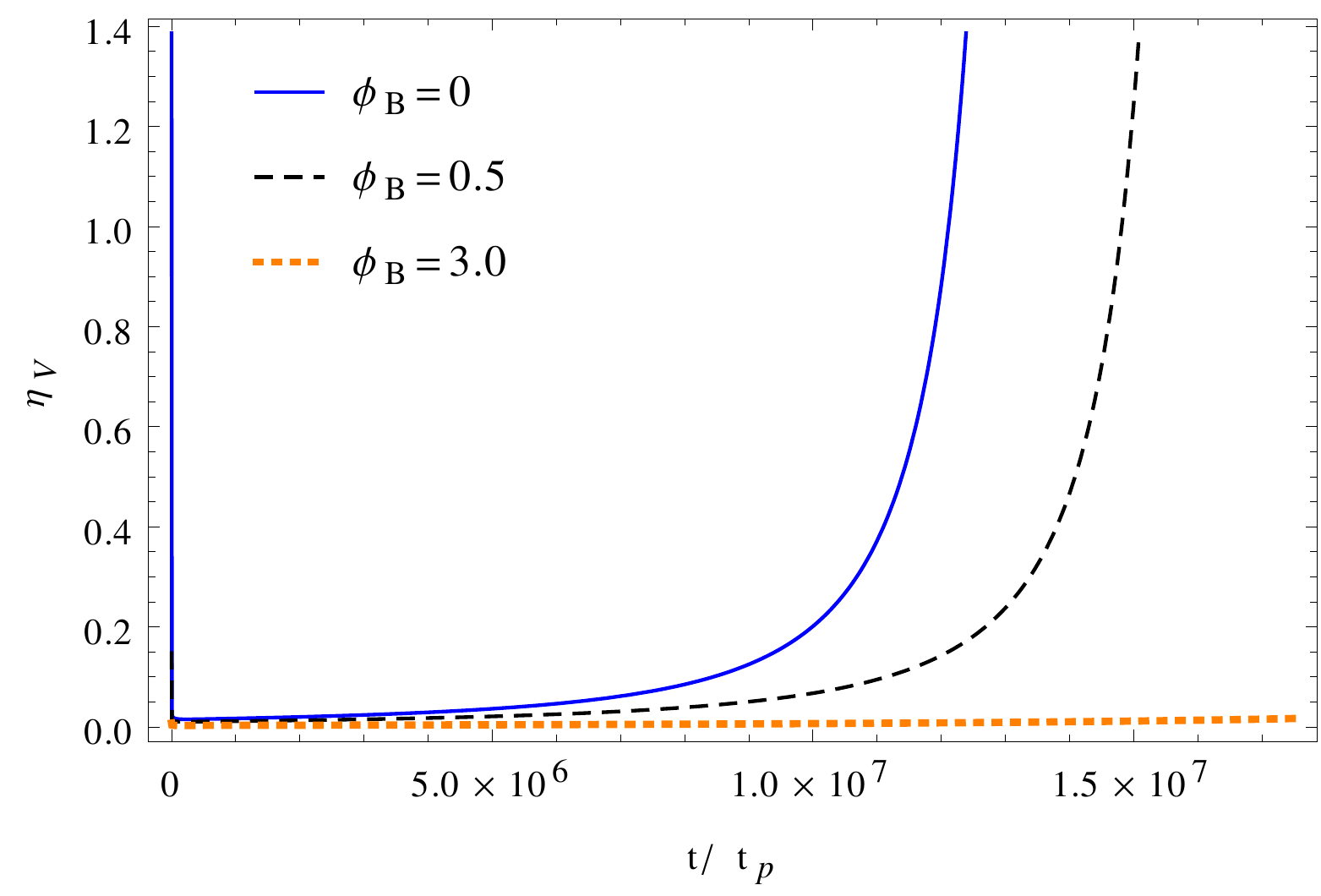} \\
   
 \end{array}
 $
  
  \caption{The \emph{left plot} is for $\epsilon_V$ and the \emph{right plot} for absolute value of $\eta_V$ with $b=0.001$, $f=0.1$ and $\delta=0$ for quadratic monodromy potential of the form $V(\phi)= \mu^3 [\frac{\phi^2}{\mu} + bf cos(\frac{\phi}{f} +\delta)]$ with $\mu = 0.749533 \times 10^{-6} m_{Pl} $ and setting $m_{Pl}=1$.}
 \label{Fig4} 
\end{figure}
In this subsection, following the previous subsection \ref{LinearMonodromy}, we use the  Friedmann Eq.(\ref{FD}) and the continuity Eq.(\ref{KG}) to numerically simulate the background dynamics, now, for a potential dominated by a quadratic monodromy term and a modulation. All the figures produced are for initially kinetically dominated universe and w.r.t cosmic time expressed in units of Planck time. In figures \ref{Fig3} and \ref{Fig4} we show the relevant dynamical quantities of the background with monodromy potential of the form $V(\phi)= \mu^3 [\frac{\phi^2}{\mu} + bf cos(\frac{\phi}{f} +\delta)]$ with values of parameters $(b,f,\delta)=(0.001,0.1,0)$ for different initial conditions of $\phi_B.$ In the left of Fig.\ref{Fig1} we plot the natural logarithmic of the scale factor $a(t)$ and on the right we plot the equation of state parameter $w_{\phi}$. The explicit expression of $w_{\phi}$ for a quadratic monodromy potential with a modulation term looks like
\begin{equation}
w(\phi) = \frac{\frac{\dot{\phi}^2}{2}-\mu^3\left[\frac{\phi^2}{\mu} + bf cos\left( \frac{\phi}{f} +\delta\right)\right]}{\frac{\dot{\phi}^2}{2} + \mu^3\left[ \frac{\phi^2}{\mu} + bf cos\left( \frac{\phi}{f} +\delta\right)\right]}.
\label{wphiQuadratic}\end{equation}
Coming to the analysis of Fig.\ref{Fig3}, the parameter $w(\phi)$ starts with almost $1$ and then slowly transits to $-1.$ This can be clearly seen from the expression above Eq.(\ref{wphiQuadratic}). For small value of $\phi_B$ the potential term in both the numerator and denominator of Eq.(\ref{wphiQuadratic}) can be neglected in comparison to the kinetic term calculated using the Eq.(\ref{BC1}). This smallness of the value of the potential term near $\phi_B$ close to zero is possible because the product $bf$ is always set to less than unity. Thus the equation of state parameter becomes $w_{\phi}\simeq 1$ initially and hence acts like stiff fluid. As the universe evolves it transits from $w_{\phi}\simeq 1$ to $\simeq - 1$ which represents the slowly rolling phase of the inflation to be verified below.  As in the case of linear monodromy, this can be easily understood from Eq.(\ref{wphiQuadratic}). As the potential energy dominates over the kinetic energy $w_{\phi}$ becomes $-1.$ No matter with what  initial value of $\phi_B$ we start out, the universe gets attracted to the $w_{\phi}=-1$ solution for the KED cases. As stated above the different values of initial conditions $\phi_B$ will effect the duration of the inflationary phase to be seen below when we calculate the number of e-folds in Tables \ref{table4} and \ref{table5}. The three phases of evolution, namely, bouncing, transition and inflation is noted here also. The behavior of the log of scale factor $a(t)$ depicts the universality of the solutions in the bouncing phase for kinetic energy dominant initial conditions at the quantum bounce. 

Now coming to the analysis of  Fig.\ref{Fig4}, the two slowly rolling parameters, $\epsilon_V$ and the absolute value of $\eta_V$  are plotted. A slow roll inflation is guaranteed if $|\eta_V|$,$\epsilon_V<<1.$ And Fig.\ref{Fig4} shows the satisfaction of these two conditions. From Eq.(\ref{SRP}) it can be easily seen that when the potential energy dominates and varies very slowly w.r.t the scalar field $\phi$, the values of $\epsilon_V,$ $|\eta_V|<<1$ and Fig.\ref{Fig4} depicts the same. The condition of slow roll inflation in this regime is satisfied untill and unless $\epsilon_V=1.$ The plots in Fig.\ref{Fig4} are to show, explicitly, the slow roll regime of the evolution of the universe.

Next we focus on finding the critical value of $r_w^c$ for quadratic monodromy potential. In order to do this we choose a set values of parameters $(b,f,\delta)=(0.001,0.1,0)$ and see for what value of $\phi_B$ leads to the generation of nearly $60$ e-folds of inflation. We found that the value of $\phi_B$ which approximately gives this number is $0.775$ (in units of $m_{Pl}$)  which gives the value of $r_w^c=1.28029 \times 10^{12}.$ The results are listed in Table \ref{table4}. Next we list the effects of variation of the product $bf$ on $N_{inf}$ in Table \ref{table5}. To our observation, in Table \ref{table5}, we notice that there is no significant change in the value of $N_{inf}$ for quadratic monodromy case unlike that of linear monodromy case as listed in Table \ref{table2}. Also, in our calculation not shown here to avoid repetition, we observed that there is even no significant change in the value of $N_{inf}$ upon the variation of the phase factor $\delta$ of the modulation term. We attribute this observation to the fact that the background dynamics is heavily dominated by the quadratic monodromy term in comparison to the modulation term unlike that of the linear monodromy case.

\begin{table}

\begin{center}

\begin{tabular}{ccccc}
\hline\hline
 $\phi_B$~~~  & Inflation~~~ & $t/t_{pl}$~~~ & $\phi_{*}$~~~ & $N_{inf}$ \\
\hline\hline
\\
0 ~~~&starts~~~& $7.8013 \times 10^4$ ~~~& 2.221~~~ & 33.973 \\
 ~~~&ends~~~& $1.191 \times 10^7$ ~~~& 0.3135~~~ &  \\\\
 
\\
0.001 ~~~&starts~~~& $7.984 \times 10^4$ ~~~& 2.223~~~ & 34.298 \\
 ~~~&ends~~~& $1.187 \times 10^7$ ~~~& 0.3232~~~ &  \\\\
 
 \\
0.1 ~~~&starts~~~& $7.4915 \times 10^4$ ~~~& 2.315~~~ & 36.776 \\
 ~~~&ends~~~& $1.247 \times 10^7$ ~~~& 0.3097~~~ &  \\\\
 
 \\
0.5 ~~~&starts~~~& $6.4583 \times 10^4$ ~~~& 2.693~~~ & 49.055 \\
 ~~~&ends~~~& $1.458 \times 10^7$ ~~~& 0.3207~~~ &  \\\\
 
 \\
0.755 ~~~&starts~~~& $5.9333 \times 10^4$ ~~~& 2.93~~~ & 59.92 \\
 ~~~&ends~~~& $2.194 \times 10^7$ ~~~& 0.02895~~~ &  \\\\
 
 \\
1.0 ~~~&starts~~~& $5.508 \times 10^4$ ~~~& 3.169~~~ & 67.127 \\
 ~~~&ends~~~& $1.737 \times 10^7$ ~~~& 0.3114~~~ &  \\\\
 
 \\
1.5 ~~~&starts~~~& $4.7875 \times 10^4$ ~~~& 3.621~~~ & 88.026 \\
 ~~~&ends~~~& $2.007 \times 10^7$ ~~~& 0.3211~~~ &  \\\\
 
 \\
2.0 ~~~&starts~~~& $4.2341 \times 10^4$ ~~~& 4.104~~~ & 112.713 \\
 ~~~&ends~~~& $2.284 \times 10^7$ ~~~& 0.3213~~~ &  \\\\
 
 \\
3.0 ~~~&starts~~~& $3.4364 \times 10^4$ ~~~& 5.076~~~ & 169.881 \\
 ~~~&ends~~~& $2.838 \times 10^7$ ~~~& 1.094~~~ &  \\\\

\hline\hline
\end{tabular}

\end{center}
\caption{This table shows the values of the number of e-folds, $N_{inf}$, for various initial values $\phi_B$ with fixed $b=0.001$ and $f=0.1$ for potential of the form $V(\phi)= \mu^3\left[ \frac{\phi^2}{\mu} + bf cos\left(\frac{\phi}{f}+ \delta\right)\right]$ with $\delta=0$ and $\mu=0.749533 \times 10^{-6}m_{Pl}$. }\label{table4}
\end{table}
\begin{table}

\begin{center}

\begin{tabular}{cccc}
\hline\hline
 $\phi_B$~~~  & $(b,f)=(0.001,0.1)$~~~ & $(b,f)=(0.095,0.79)$~~~ & $(b,f)=(0.999,0.999)$ \\
\hline\hline
\\
$0$ ~~~&$N_{inf}=33.963$~~~& $N_{inf}=35.939$ ~~~& $N_{inf}=35.613$ \\
 ~~~&$t_i= 7.8013 \times 10^4$~~~& $t_i= 7.8017 \times 10^4$ ~~~& $t_i= 7.801 \times 10^4$ \\
  ~~~&$t_f= 1.191 \times 10^7$~~~& $t_f= 1.188 \times 10^7$ ~~~& $t_f= 1.188 \times 10^7$
  \\\\

$0.755$ ~~~&$N_{inf}=59.92$~~~& $N_{inf}=59.92$ ~~~& $N_{inf}=59.964$ \\
 ~~~&$t_i= 5.9333 \times 10^4$~~~& $t_i= 5.9333 \times 10^4$ ~~~& $t_i= 5.9333  \times 10^4$ \\
  ~~~&$t_f= 2.194 \times 10^7$~~~& $t_f= 2.194 \times 10^7$ ~~~& $t_f= 2.194 \times 10^7$
  \\\\
  
$3.0$ ~~~&$N_{inf}=169.881$~~~& $N_{inf}=169.881$ ~~~& $N_{inf}=169.881$ \\
 ~~~&$t_i= 3.4364 \times 10^4$~~~& $t_i= 3.4364 \times 10^4$ ~~~& $t_i= 3.4365 \times 10^4$ \\
  ~~~&$t_f= 2.838 \times 10^7$~~~& $t_f= 2.84 \times 10^7$ ~~~& $t_f= 2.48 \times 10^7$
  \\\\
\hline\hline
\end{tabular}

\end{center}
\caption{This table shows the values of the number of e-folds, $N_{inf}$, w.r.t. the variation of the product of the parameters $bf$ for three different initial values $\phi_B$ for potential of the form $V(\phi)= \mu^3\left[ \frac{\phi^2}{\mu} + bf cos\left(\frac{\phi}{f}+ \delta\right)\right]$ with $\delta=0$ and $\mu=0.749533 \times 10^{-6}m_{Pl}$.}\label{table5}
\end{table}

\section{Dynamical System Analysis}\label{DySA}

A dynamical system analysis gives a qualitative behavior of the system in question. It is well known that the background dynamics of cosmology can be formulated and treated as a dynamical system \cite{WE}. The popular method, to carry out dynamical system analysis, followed in the community of cosmology is the Linear Stability Analysis \cite{DSA1}. It involves formulation of the problem in terms of first order autonomous differential equation and the flow vector of the system  is used to determine the qualitative behaviour. For the readers interested in alternative methods we refer them to \cite{KCC,ChenZhu}.

The phase portrait  of a dynamical system depicts a pictorial and geometric representation of the trajectories of the system. It gives us a qualitative and intuitive idea for the fate of  trajectories by drawing the tangents to the trajectories at each point dictated by the flow vector. In the following we discuss the phase portraits due to two different monodromy potential considered in the Sec.\ref{Cosmology} for an FLRW universe in the framework of LQC.

\subsection{Dynamical System Analysis for $V(\phi)= \mu^3\left[\phi+bf cos\left(\frac{\phi}{f}+\delta\right)\right]$}\label{DySA1}

\subsubsection{Phase portrait}\label{PP1}
The key to drawing the phase portrait is to reformulate the problem in terms of first order differential equations. This requires the introduction of normalized, often dimensionless, dynamical variables and then taking the derivative w.r.t. the newly defined time to arrive at the first order differential equation. There could be numerous ways to define the variables. The choice entirely depends on way the we want to represent the phase-portrait. Let us begin with the energy density of the scalar field which is given by
\begin{equation}
\rho = \frac{{\dot{\phi}}^2}{2} + V(\phi).
\end{equation}
Substituting the linear monodromy potential, it becomes
\begin{eqnarray}\nonumber
\rho &=&  \frac{\dot{\phi}^2}{2} + {\mu}^3 \left[ \phi + bf cos \left(\frac{\phi}{f}+ \delta\right)\right], \\
      &=& \rho_c \left[ \left(\frac{\dot{\phi}}{\sqrt{2\rho_c}}\right)^2 + \frac{\mu^3 f}{ \rho_c}\left( \frac{\phi}{f}+ bcos\left(\frac{\phi}{f}+\delta\right)\right)\right]. \label{EE1}
\end{eqnarray}
Now let us define a new set of dimensionless dynamical variables as $x \equiv \frac{\phi}{f}$ and $y \equiv \frac{\dot{\phi}}{\sqrt{2 \rho_c}}$. With these choice of variables Eq.(\ref{EE1}) becomes
\begin{equation}
\frac{\rho}{\rho_c}= y^2+ \frac{\mu^3 f}{\rho_c}\left[x+bcos(x+\delta)\right]. 
\label{Constraint1}
\end{equation}
Thus, having defined the dynamical variables, it is now a straight forward task to get the autonomous system of dynamical equation by simply taking the time derivative of the dynamical variables to obtain the following:
\begin{eqnarray} \nonumber
\frac{dx}{dt} &=& \frac{\sqrt{2 \rho_c}}{f} y, \\
\frac{dy}{dt} &=& -3H(t)y - \frac{\mu^3}{\sqrt{2 \rho_c}}\left[ 1-bsin(x+\delta) \right]; \label{Flow1}
\end{eqnarray}
where the Klien Gordon Eq.(\ref{KG}) is used to obtain the equation for $y$. It is to be noted that $H$ in the above equation has to substituted in terms of $(x,y)$ which is
\begin{equation} \nonumber
 H(t)=\sqrt{\frac{8 \pi G}{3} \rho_c \left[ y^2 + \frac{\mu^3 f}{\rho_c}\left( x + b cos\left(x+ \delta\right)\right)\right]\left[ 1- \left( y^2 + \frac{\mu^3 f}{\rho_c}\left(x + b cos \left(x+ \delta\right)\right)\right)\right]}.\nonumber 
 \end{equation}
Thus the pair $(x,y)$ constitute the generalized coordinates of a two dimensional system and hence forms the phase-space of the same whose flow vector is given by the Eq.(\ref{Flow1}). 

\begin{figure}
$
 \begin{array}{c c}
   \includegraphics[width=0.45\textwidth]{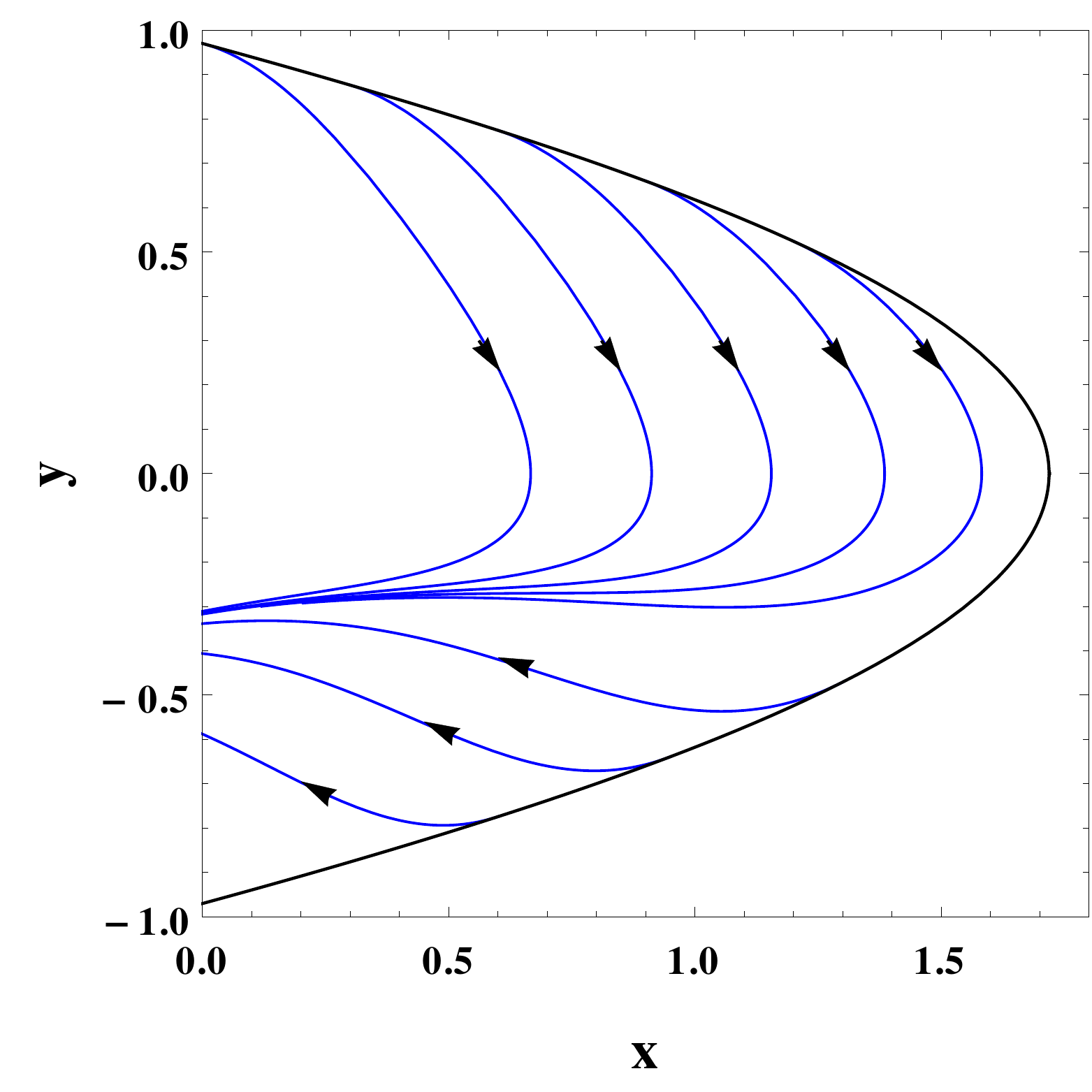} &
   \includegraphics[width=0.45\textwidth]{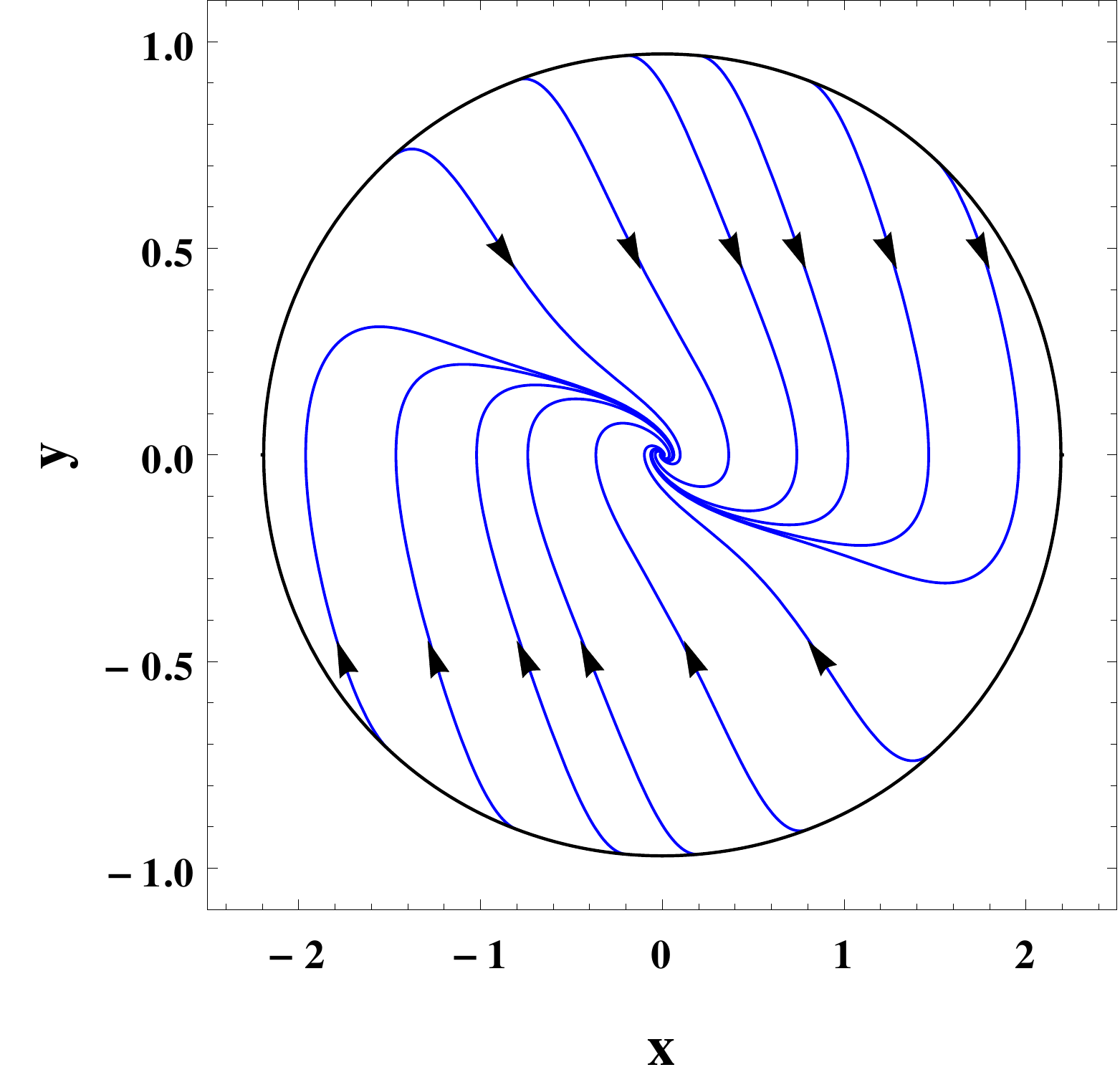} \\
   
 \end{array}
 $
  
  \caption{The \emph{left} phase portrait is for the case of linear monodromy potential and the \emph{right} for the quadratic monodromy potential with  the parameters $b=0.1$, $f=0.33$, $\delta=0$ and $\mu=0.9$ in the dynamical variable $(x, y)$.}
 \label{Phase1} 
\end{figure}

However note that, though, the phase space coordinates $(x,y)$ are dimensionless, Eq.(\ref{Flow1}) is not because of the dimension full quantity, the cosmic time $t$. It is convenient to express the equations in dimensionless time variable defined as $\tau \equiv \frac{\sqrt{2 \rho_c}}{f}t$, which reads as:
\begin{eqnarray}\nonumber
\frac{dx}{d\tau} &=& y, \\
\frac{dy}{d\tau} &=& 3 y\mathcal{H}(\tau) - \frac{f {\mu}^3}{2 {\rho_c}}\left[1 -b sin(x+\delta) \right];\label{Flow1PhasePortrait}
\end{eqnarray}
where $\mathcal{H}(\tau) \equiv \frac{1}{a(\tau)} \frac{d a(\tau)}{d \tau} = \frac{1}{a(t)}\frac{dt}{d\tau} \frac{d a}{d t} =  \frac{f}{\sqrt{2\rho_c}} H(t)$. This completes our task of making both sides of the equation dimensionless. The phase portrait in terms of the dynamical variables $(x,y)$ in $\tau$ frame are shown on the left of Fig.(\ref{Phase1}) using Eq.(\ref{Flow1PhasePortrait}). From the discussion in Sec.\ref{Cosmology}, we recall that the energy density reaches its maximum value $\rho=\rho_c$ which is at the bounce. Thus, the locus of all the points for which $\rho=\rho_c$ is satisfied constitute the boundary of the physical phase space. Because it is the inner region defined by this boundary which represents the expanding branch of the universe signified by positive value of $H$ that is of cosmological interest to us. The region outside the boundary gives the contracting universe. In fact, from cosmology point of view,  we are interested in only half of the phase space  $x \geq 0$ as we want the potential to be positive as discussed in Sec.\ref{Cosmology}. It is clear that the boundary of the portrait is certainly not a circle because of the constraint Eq.(\ref{Constraint1}) and more precisely for the choice of our dynamical variables.



Coming to the analysis of the phase diagram, the left diagram of Fig.\ref{Phase1} shows the phase portrait of linear monodromy potential with a modulation term in $(x,y)$ dynamical pair in $\tau$ frame, where we use Eq.(\ref{Flow1PhasePortrait}). Here we have used the values of parameters $(b,f)=(0.1,0.33)$ and $\mu= 0.9$ for the sake of better visibility. Note that the dark black boundary line in the figure defined by $x=0$ is certainly not the boundary of the physical phase space. This range is so fixed only to show the region of positive valued potential. In fact, the phase portrait of linear monodromy potential is not bounded when we do not restrict the dynamics to positive valued potential alone. The locus of the points where the bounce occurs is certainly $y^2+ \frac{\mu^2 f}{\rho_c}\left[ x + b cos(x+\delta) \right]=1.$ This immediately gives $y\leq \pm \sqrt{1- \frac{\mu^3 f}{\rho_c}\left[ x+ bcos(x+\delta)\right]}$ as the region wherein the universe is expanding after passing through the quantum bounce. From this, it is clear that there is an upper bound to the value of $x$ in the physical phase space. This is because there is a maximum value that $x$ can acquire for the $y$ to be a real quantity. However, there is no limit on the negative value of $x$ for the real valued physical phase space. Thus, $x$ bounded from above but not below makes $y$ unbounded. This makes the phase space non compact. Also, as said above, since in the realistic background dynamics discussed in Sec.\ref{Cosmology} we are interested only in the positive value of potential and hence $x>0.$ Therefore it is the region of $x>0$ which is of cosmological interest to us and is presented here. The analysis of the ultimate fate of the trajectories, in this case, is beyond the scope of this paper. It is clear that the phase space is not symmetric w.r.t the transformation $x\rightarrow -x$ owing to the form of the potential. Looking at the phase portrait it seems that the trajectories coming off from the bouncing eventually get attracted to a particular solution in the forward direction of time. This solution is a slowly rolling inflationary solution and as shown in \cite{NegativeCosmologyParam} which acts as an attractor in the forward direction of time and, obviously, a repellor if we reverse the time direction. Also, this solution acts as a separatrix meaning that the trajectories from one region do not go into the other region crossing the inflationary solution. Also, the visual in the left of Fig.\ref{Phase1} there is no fixed point for the trajectories to settle down. This is to be proved in subsection \ref{FiP1}. Though the occurrence of the fixed points depends heavily on the shape of the potential considered, but it should not be taken for granted and a thorough fixed point analysis must be carried out.

\subsubsection{Fixed Point Analysis}\label{FiP1}
Fixed points are the equilibrium points of the autonomous system. They are the points on the phase space where the flow vector vanishes. In this subsection we rigorously find if there is any such point and comment on their stability properties. The system of equation, from subsection \ref{PP1},  we consider for the fixed point analysis is given below:

\begin{eqnarray}\nonumber
\frac{dx}{d\tau} &=& y, \\
\frac{dy}{d\tau} &=& -3 y\mathcal{H}(\tau) - \frac{f {\mu}^3}{2 {\rho_c}}\left[1 -b sin(x+\delta) \right].
\end{eqnarray}
Let us denote the slope of the dynamical variable $x$ and $y$ as the functions $f(x,y)$ and $g(x,y)$ respectively. This gives
\begin{eqnarray}\nonumber
f(x,y) &\equiv & y, \\
g(x,y) &\equiv & 3 y\mathcal{H}(\tau) - \frac{f {\mu}^3}{2 {\rho_c}}\left[1 -b sin(x+\delta) \right].
\end{eqnarray}
The fixed points, if any, of this system are the simultaneous solutions of 
\begin{eqnarray}\nonumber
f(x,y) &\equiv & y  =0,\\
g(x,y) &\equiv &- 3 y\mathcal{H}(\tau) - \frac{f {\mu}^3}{2 {\rho_c}}\left[1 -b sin(x+\delta) \right] =0.
\end{eqnarray}
Let us denote the fixed points as $(x_c, y_c)$. It is straight forward to see that the zero is only allowed value for $y_c.$ Therefore the set of points ${(x_c,0)}$ are the points for which 
\begin{equation}
g(x_c,0)=0.
\end{equation}
This immediately gives:
\begin{equation}
x_c= arcsin\left(\frac{1}{b}\right)+ 2k\pi,
\end{equation}
where $k=(0,\pm 1, \pm 2, ..)$. Thus the set of points $(x_c, y_c)= (arcsin(\frac{1}{b})+2k\pi, 0)$ constitute the fixed point of the system. But for meaningful cosmology the parameter $b<1$ implying $\frac{1}{b}>1$. As the range of the sinusoidal function is bounded $(-1,1)$, therefore, there is no solution to $g(x_c,0)=0$ and hence no fixed point for the kind of cosmology we are interested in. Having said that, nonetheless, we present below the stability criteria for the sake of completeness for interested readers and more importantly to be used in the subsection \ref{FiP2}.

The stability property of a fixed point is determined by the fate of the solution when we perturb it from the fixed point. For example if the perturbation around a fixed point grows exponentially then it is an unstable fixed point. Whereas if the perturbation around a fixed point is suppressed exponentially then it is a stable fixed point. Mathematically, this is carried out by linearizing the system:

\begin{equation}
\left(\begin{array}{c}
{d (\delta {x})}/{ d\tau} \\ 
{d (\delta {y})}/{ d\tau}
\end{array}\right)  =\left(\begin{array}{cc}
\frac{\partial f(x,y)}{\partial x} & \frac{\partial f(x,y)}{\partial y} \\ 
\frac{\partial g(x,y)}{\partial x} & \frac{\partial g(x,y)}{\partial y}
\end{array} \right)_{(x_c,y_c)} \left(\begin{array}{c}
\delta {x} \\ 
\delta {y}
\end{array}\right), 
\label{Linearized}
\end{equation}
where $\delta {x}$ and $\delta {y}$ represent small perturbation around the fixed point $(x_c,y_c)$. Denoting the coefficient matrix of the above Eq.(\ref{Linearized})as $\textbf{A}$, the properties of a fixed point can be determined by looking at the nature of the eigenvalues of diagonalized matrix $\text{A}$. This amounts to find the roots of the quadratic equation $|\textbf{A}-\lambda \textbf{I}|=\textbf{0}$ which with $\delta=0$ gives:
\begin{equation} \label{Char}
{\lambda^2} + 3\lambda \mathcal{H}(x_c,y_c) - \alpha cos(x_c)=0,
\end{equation}
where $\alpha\equiv \frac{f \mu^3 b}{2 \rho_c}$. The roots of this equation are
\begin{equation} \label{Roots}
\lambda_{1,2} = \frac{-3\mathcal{H}(x_c,y_c) \pm \sqrt{9\mathcal{H}^2(x_c,y_c)+ 4 \alpha cos(x_c)}}{2}.
\end{equation}
The stability criteria can be classified according to the signs of the eigenvalues $\lambda_1$ and $\lambda_2.$ The types of fixed points can be broadly divided into three catogories depending upon the sign of the discriminant which for our case is $D\equiv 9 \mathcal{H}^2(x_c,y_c)+4\alpha cos(x_c)$. These are stated as below. 

\textbf{\textit{Case 1:)}$D>0$} For this both $\lambda_1$ and $\lambda_2$ are real.
This implies that for a given fixed point $(x_c,y_c)$ the region of parameter space must satisfy the following relation:
\begin{equation}
D\equiv 9 \mathcal{H}^2(x_c,y_c)+4\alpha cos(x_c)>0.
\label{RPM1}
\end{equation}
In this case we will have two distinct eigenvalue namely $\lambda_1= -3\mathcal{H}(x_c,y_c)-\sqrt{D}$ and $\lambda_2= -3\mathcal{H}(x_c,y_c)+\sqrt{D}$. It is straightforward to see that $\sqrt{D}$ will be either greater or less than $|-3\mathcal{H}(x_c,y_c)|$ depending on whether $cos(x_c,y_c)$ assumes positive or negative value respectively. In case $\sqrt{D}<|-3\mathcal{H}(x_c,y_c)|$ then both the eigenvalues $\lambda_{1,2}$ will come with negative sign. This would give rise to stable node. Now in case of $\sqrt{D}>|-3\mathcal{H}(x_c,y_c)|$ would lead to eigenvalues with opposite signs. The eigenvalues with opposite signs signifies unstable fixed point or saddle point singularity.
 
\textbf{\textit{Case 2:)}$D<0$} Both $\lambda_1, \lambda_2$ are complex, i.e. of the form $\lambda_{1,2}= w\pm iz$. For this to be true we must have $9 \mathcal{H}^2(x_c,y_c)+4\alpha cos(x_c)<0$. Two distinct cases may arise here. One in which $w<0$ and the other $w>0$. 

For $w\neq0$ the fixed point is a spiral or a focus, that is, the solutions approach the fixed point as $t \rightarrow \infty$  but not from a definite direction. This can be, further, divided into two cases: $w<0$ and $w>0$. For $w<0$ the fixed point is a stable focus whereas for $w>0$ it is an unstable focus. Since we are dealing with an expanding phase of the universe, therefore, the Hubble parameter is positive and hence $w<0$ for the fixed point $(x_c, y_c)$. We will see this explicitly in the subsection \ref{FiP2} for quadratic monodromy potential case. For $w=0$ implies the real parts of the eigenvalues are zero which give rise to non hyperbolic fixed points. The fixed point is a center, that is not stable in the usual sense and therefore one must look at higher order derivatives to check their stability.
 
\textbf{\textit{Case 3:)}$D=0$} For this $\lambda_1$ and $\lambda_2$ are equal. In this case if there are two linearly independent eigenvectors we have a star singularity whereas we have an improper node if there is only one linearly independent eigenvactor.

\subsection{Dynamical System analysis for $V(\phi)= \mu^3\left[\frac{\phi^2}{\mu}+bf cos\left(\frac{\phi}{f}+\delta\right)\right]$}\label{DySA2}

\subsubsection{Phase Portrait}\label{PP2}

This subsection  studies the phase portrait of the background dynamics for a universe dominated by a quadratic monodromy term with a small modulation in LQC set-up. Here we follow exactly, the procedure mentioned in the previous section for linear monodromy potential in subsection \ref{PP1}. Therefore, we begin with the energy density of the scalar field which for our choice of potential becomes:
\begin{eqnarray}\nonumber
\rho &=& \frac{\dot{\phi}^2}{2} + \mu^3 \left[ \frac{\phi^2}{\mu} + b f cos \left(\frac{\phi}{f}+ \delta\right)\right], \\
 &=& \rho_c \left[  \left( \frac{\dot{\phi}}{\sqrt{2 \rho_c}}\right)^2 +  \frac{\mu^3 f}{\rho_c} \left( \frac{\phi^2}{f \mu} + b cos\left(\frac{\phi}{f}+ \delta\right)\right) \right].\label{EE2}
\end{eqnarray}
Defining  the dimensionless dynamical variable as that in subsection \ref{PP1}: $x\equiv\frac{\phi}{f}$ and $y\equiv \frac{\dot{\phi}}{\sqrt{2 \rho_c}}$ we have
\begin{equation}\label{Constraint2}
\frac{\rho}{\rho_c}= y^2 + \frac{\mu^3 f}{\rho_c} \left[ \frac{f x^2}{\mu} + b cos(x+ \delta) \right]. 
\end{equation}
Taking the time derivative of the above defined dynamical variables $x$ and $y$ and using the Klien Gordon equation for $y$, we have
\begin{eqnarray}\nonumber
\frac{dx}{dt} &=& \frac{\sqrt{2 \rho_c}}{f} y,\\
\frac{dy}{dt} &=& -3H(t)y - \frac{\mu^3}{\sqrt{2 \rho_c}}\left[ 2 \left(\frac{f}{u}\right)x - b sin(x+ \delta) \right];\label{Flow2}
\end{eqnarray}
where $H(t)=\sqrt{\frac{8 \pi G}{3} \rho_c \left[y^2 + \frac{\mu^3 f}{\rho_c} \left( \frac{f}{\mu} x^2 + bcos(x+\delta) \right) \right] \left[ 1- y^2 -\frac{\mu^3 f}{\rho_c}\left( \frac{f}{\mu}x^2 + b cos(x+\delta)\right)\right]}$. Now, introducing the dimensionless time $\tau= \frac{\sqrt{2\rho_c}}{f} t$ and expressing the above set of equation in $\tau$ frame we get: 
\begin{eqnarray}\nonumber
\frac{dx}{d\tau} &=& y, \\
\frac{dy}{d\tau} &=& -3\mathcal{H}(\tau) y - \frac{f \mu^3}{2 \rho_c}\left[  2 \left(\frac{f}{\mu}\right) x - b sin(x+\delta)\right]; \label{Flow2PhasePortrait}
\end{eqnarray} 
where $\mathcal{H}(\tau)=\frac{dt}{d\tau}H(t)$. As can be seen from the Eq.(\ref{Constraint2}) the locus of initial condition defined by $\rho=\rho_c$ is definitely not a circle in the dynamical pair of variable $(x,y)$. Eq.(\ref{Flow2PhasePortrait}) is used to produce the right hand figure in Fig.\ref{Phase1}.

Coming to the analysis of the phase portrait, right diagram of the Fig.\ref{Phase1} shows the phase portrait of a quadratic monodromy potential with a modulation term in $(x,y)$ dynamical pair of variables in $\tau$ frame. The boundary of the physical phase space is the locus of all the points for which $y^2+ \frac{\mu^3 f}{\rho_c}\left[ \frac{f}{\mu} x^2 + b cos(x+\delta)\right]=1$ is satisfied. Hence the physical region of phase space is given by the inequality  $y^2+ \frac{mu^3 f}{\rho_c}\left[ \frac{f}{\mu} x^2 + b cos(x+\delta)\right] \leq 1$ and the range of $y$ is: $|y| \leq  \sqrt{ 1 - \frac{\mu^3 f}{\rho_c}\left[  \frac{f}{\mu}x^2 + b cos(x+\delta)\right]}.$ It is clear that for $y$ to be real the value of $x$ must be bounded both in the positive and negative direction. This implies that the range of $y$ is also bounded from both above and below. Here the region of physical space unlike that of linear monodromy case is compact. The boundary shown by the dark black curve in the right of Fig.\ref{Phase1} is actually the boundary of the physical phase space signifying the expanding branch which is not perfectly circular because of the small modulation term. Note that the trajectories tend to meet a particular solution which is the inflationary solution. The inflation serves as a separatrix which divides the phase space into two regions. The phase portrait reveals that the trajectories, in the asymptotic limit, go in spiraling towards the origin. This suggests that the origin $(0,0)$ is a spiral stable fixed point or also called stable focus of the system. A thorough fixed point and stability analysis is to be carried out in subsection \ref{FiP2} before coming to any conclusion.

\subsubsection{Fixed Point Analysis}\label{FiP2}
\begin{figure}
$
 \begin{array}{c c c c}
   \includegraphics[width=0.45\textwidth]{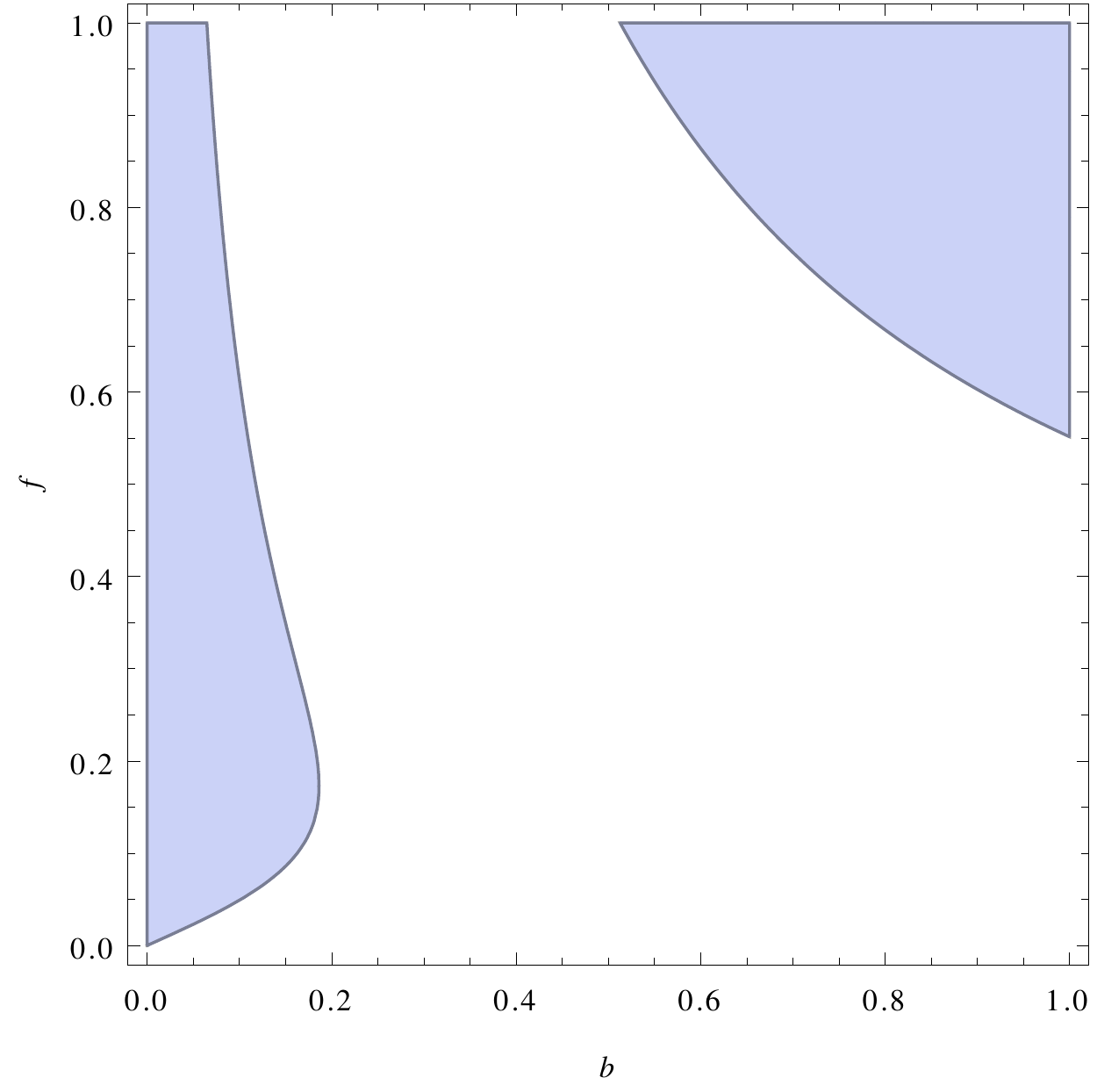} &
   \includegraphics[width=0.45\textwidth]{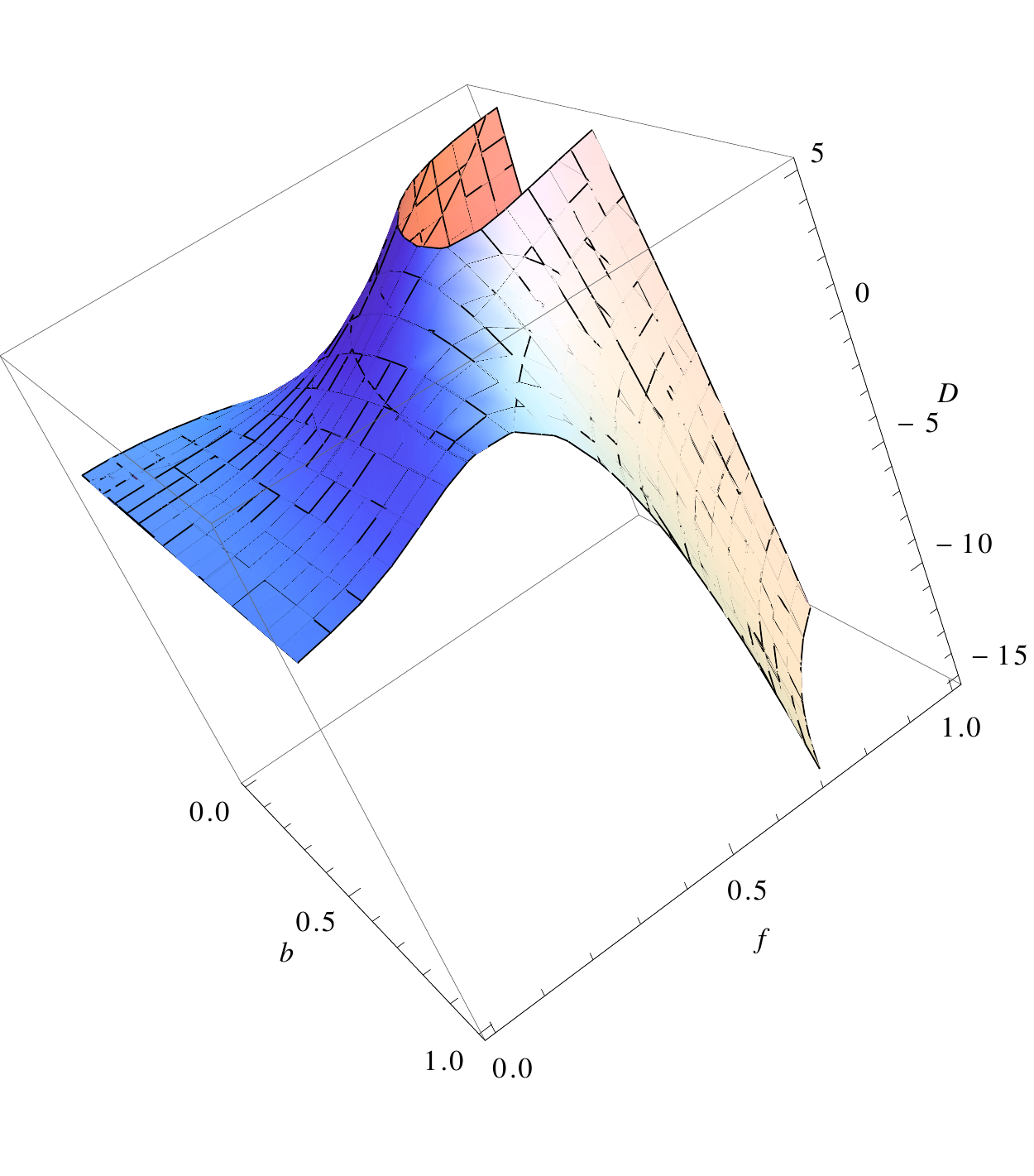} \\
  
   \includegraphics[width=0.45\textwidth]{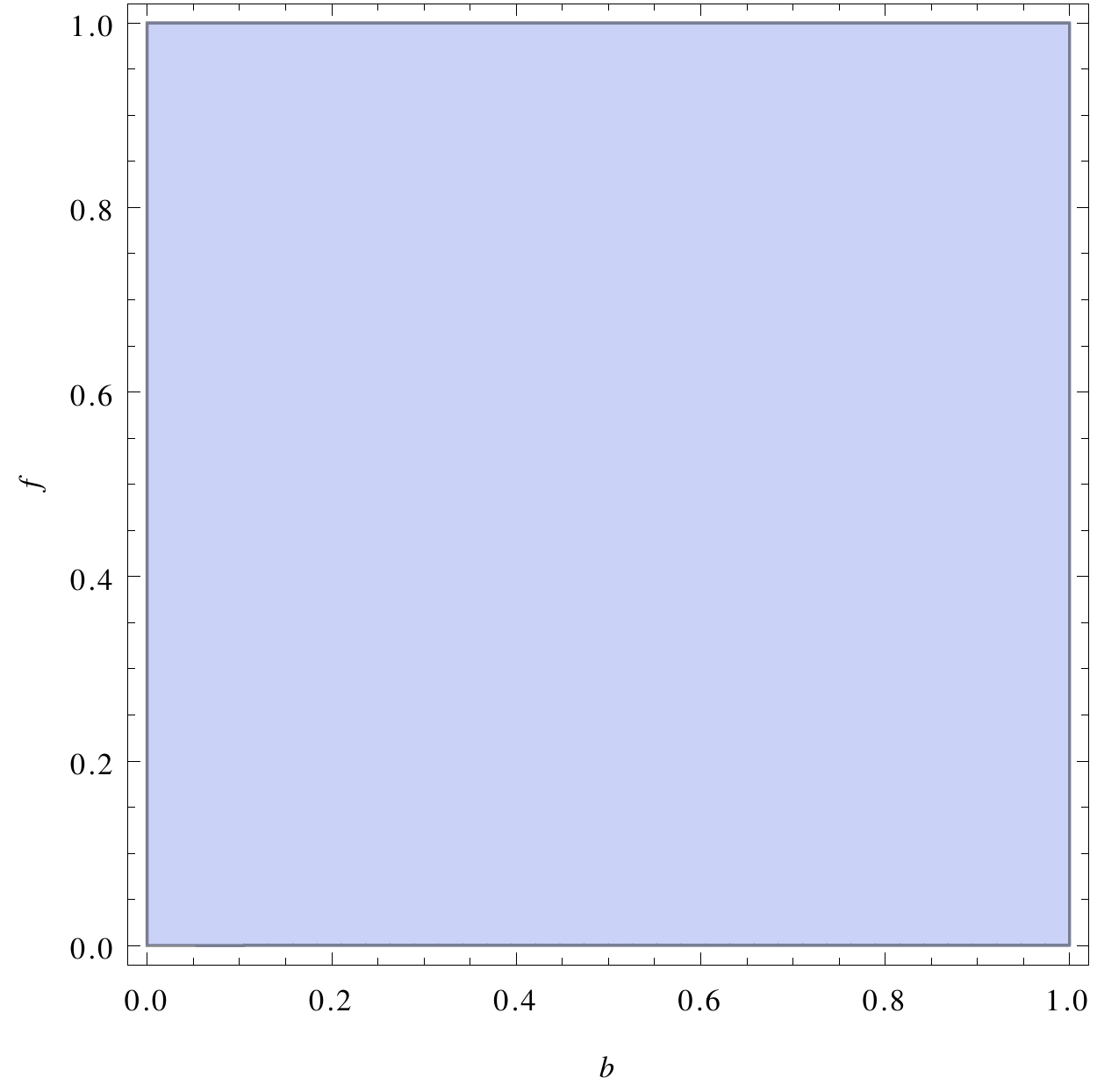} &
   \includegraphics[width=0.45\textwidth]{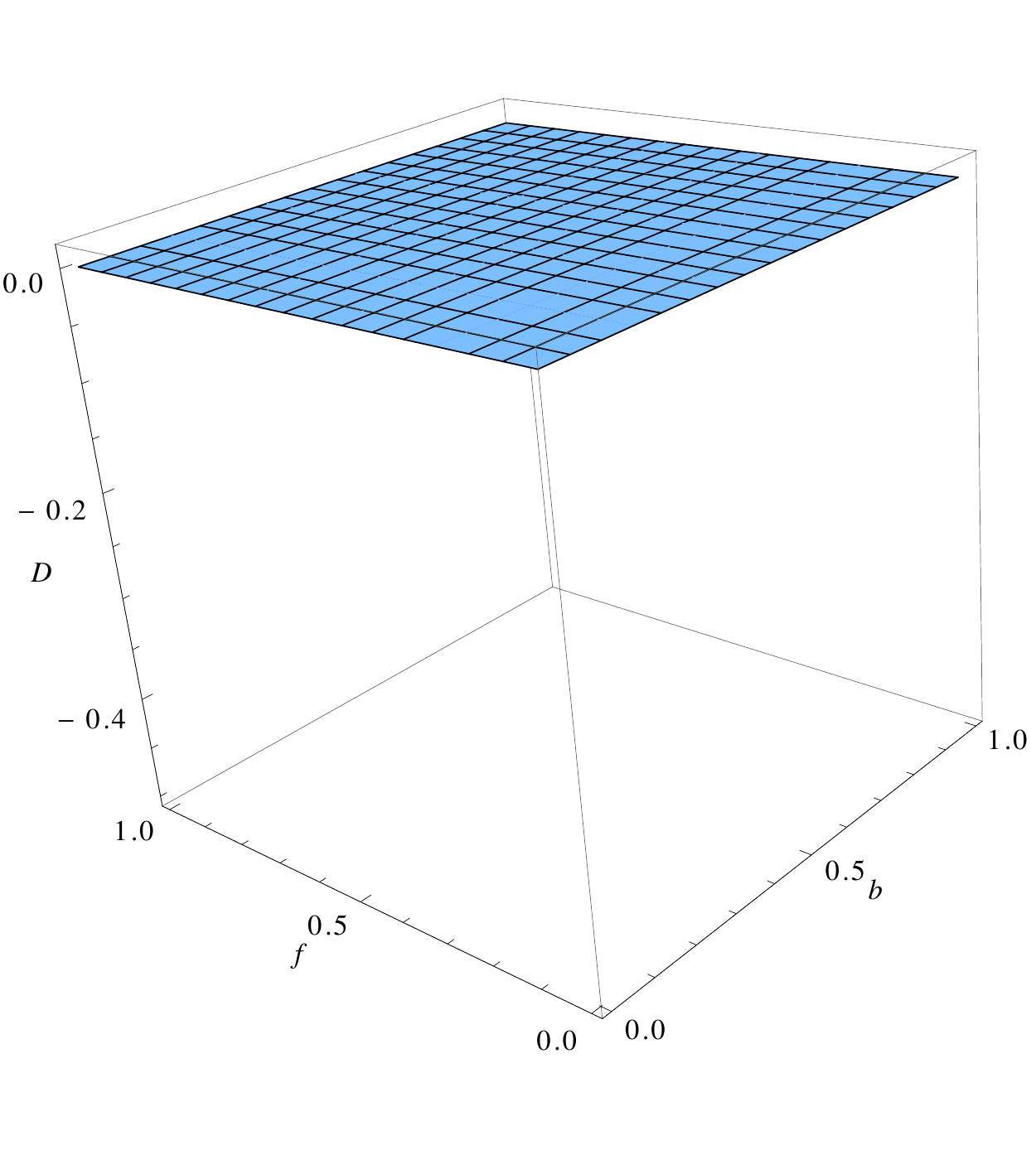} \\

 \end{array}
 $
  
  \caption{\emph{Top Left:} Region of allowed parameter space, \emph{ Top Right:} with specified$D$ for $\mu=0.9$ and \emph{Bottom Left:} Region of allowed parameter space, \emph{ Bottom Right:} with specified$D$ for $\mu=10^{-6}$.}
\label{Region}
\end{figure}

The fixed point of the dynamical system dominated by a monodromy potential with quadratic case is obtained by simultaneously solving the slopes in Eq.(\ref{Flow2PhasePortrait}) by setting them to zero. It is easy to see that the point $(x_c, y_c)=(0,0)$ is a fixed point of the system. Also from the right of the phase-portrait in Fig.\ref{Phase1} it seems that the fixed point $(0,0)$ acts as an attractor. Below we analyze this point with rigour. Linearizing about the fixed point $(x_c,y_c)=(0,0)$, the coefficient matrix \textbf{A} for the system of Eq.(\ref{Flow2PhasePortrait}) evaluated at the point $(0,0)$ is 
\begin{equation}
\textbf{A} = \left(\begin{array}{cc}
\frac{\partial f(x,y)}{\partial x} & \frac{\partial f(x,y)}{\partial y} \\ 
\frac{\partial g(x,y)}{\partial x} & \frac{\partial g(x,y)}{\partial y}
\end{array} \right)_{(0,0)} = \left(\begin{array}{cc}
0 & 1 \\ 
\overline{\alpha} &-3\mathcal{H}(0,0)
\end{array} \right)  , 
\label{Linearized1}
\end{equation}
where $\overline{\alpha}= - \frac{f \mu^3}{2 \rho_c}\left[ 2 \frac{f}{\mu} - b\right]$. The stability of the fixed point is determined by the sign of the eigenvalues obtained by diagonalizing the matrix. Solving the characteristic equation $|\textbf{A}-\lambda \textbf{I}|=0$ we get:
\begin{equation}
\lambda^2 + 3 \mathcal{H}(0, 0) \lambda -\overline{\alpha}=0,
\end{equation}
which gives
\begin{equation}
\lambda_{1,2}= \frac{- 3 \mathcal{H}(0, 0) \pm \sqrt{9 \mathcal{H}^2(0, 0)+4\overline{\alpha}}}{2}.
\end{equation}
Now let us examine the eigenvalues. The general form of the eigenvalue can be written as $\lambda_{1,2}= w \pm i z.$ Where $z\equiv \sqrt{D}$ whereas $D$ is the discriminant. In our case $D=\sqrt{9\mathcal{H}^2(0, 0) + 4 \overline{\alpha}}$. We shall examine the sign of D after a while. For the case of $D<0$ this can, further, be divided into three cases, namely, $w=0,-1$ and $1$. For $w=0$ it is called a non hyperbolic fixed point whose stability can be analyzed by going to higher order derivatives. Moreover, $w=0$ is cosmologically not achieved as $w=-3 \mathcal{H}(0,0) \neq 0.$ For $w\neq 0$ the fixed point is a spiral or a focus. This means the solutions approach the fixed point asymptotically but not from a definite direction provided $D<0$. In our case, as we are in the expanding branch of the universe, therefore, $w= -3\mathcal{H}(0,0)<0$. Now  checking the value of $D$ for the set of parameters $(b,f, \mu)=(0.1,0.33, 0.9)$, being used to produce the right diagram of the panel of Fig.\ref{Phase1}, we get $D= -0.289715$ which is negative. Whereas for $(b,f,\mu)= (0.1, 0.33, 10^{-6})$, the discriminant $D= -1.19376 \times 10^{-12}$ which is still less than zero. Note that $\mu=  1.06 \times 10^{-6}$ corresponds to the realistic background dynamics discussed in Sec.\ref{Cosmology}. Below we note the eigenvalues corresponding to two set of parameters $(b, f, \mu)=(0.1, 0.33, 0.9)$ and $(b,f,\mu)=(0.1,0.33,10^{-6})$.
The eigenvalue for $(b, f, \mu)=(0.1, 0.33, 0.9)$:
\begin{equation}
\lambda_{1,2}= \frac{-1.30669 \pm 0.53826 \textit{i}}{2},
\end{equation}
and for $(b,f,\mu)=(0.1,0.9,10^{-6})$:
\begin{equation}
\lambda_{1,2}= \frac{-1.57738 \times 10^{-9} \pm 1.0925 \times 10^{-6} \textit{i} }{2}.
\end{equation}
We also draw the region of parameter space which gives rise to a spiral or stable focus fixed point treating $b$ and $f$ as free parameters while keeping $\mu$ fixed on the top and bottom left of Fig.\ref{Region}. In addition to this, we specify the value of the corresponding discriminant $D$ by using the 3-d contour graphics on the top and bottom right of the Fig.\ref{Region} . The top panel of Fig.\ref{Region} is used for the set of parameters $(b, f, \mu)=(0.1,0.33,0.9).$ On its left we show the allowed region of parameter space for which the fixed point $(0,0)$ is spiral, while, on the right we provide the proof by specifying the value of $D$ which is negative for the allowed region and positive otherwise. Note that it is only within and on the region of allowed parameter space that it will show as spiral fixed point. The same goes for the bottom panel of the figure except that $\mu=10^{-6}.$ Note that as we decrease the value of the mass term $\mu$ of the monodromy potential, the split in the allowed region of the parameter space starts disappearing. And for the realistic background dynamics, $\mu \sim 10^{-6}$, all values of parameters within the range $(0<b<1, 0<f<1)$  are allowed. And hence, the value of $D$ for this case is a very small negative number for the whole range of the values of parameters $(b,f).$ Hence the point $(0,0)$ is a stable spiral or focus.

\section{Discussion of the result}\label{DR}
In this article we investigated the pre-inflationary background dynamics of a universe sourced by a scalar field in an FLRW background in LQC framework. We considered two different kinds of monodromy potentials, namely, linear and quadratic with a modulation term to this effect.
As the universality of the scale factor is widely noticed in various literature \cite{NewPaperA,NewPaperB} and is nor an exception for Monodromy potential which drives the inflation. Hence for a monodromy potential also, when we initialize our universe with a kinetic dominated bouncing state, it goes through three different phases respectively: \textit{bouncing, transition and slow roll}. While we use the plots of natural logarithmic of scale factor and $w_{\phi}$ to denote the full evolution of the universe starting from the bounce, the plots of slow roll parameters are used here to capture the slowly rolling phase of the universe.

A considerable space of this paper is dedicated towards understanding the effects of positive valued monodromy potential with a modulation term on the background dynamics. Here also we take these two different monodromy potentials, namely, linear and quadratic case by case. In particular, we calculate the number of e-folds of expansion generated in slowly rolling phase of inflation due to these two potentials separately. It is well known that the inflaton must roll slowly enough to produce the sufficient number of e-folds to cure the puzzles of the standard model of cosmology. In this article we take this benchmark to be nearly $60$  and see if it is generated and if so for what initial  value $\phi_B.$ In order to do this, we first set the parameters $(b,f,\delta)$ to $(0.001,0.1, 0)$ and vary the only parameter $\phi_B$ as $\dot{\phi}_B$ is already determined up to a sign due to the bouncing condition Eq.(\ref{BC1}) as discussed in Sec.\ref{Cosmology}. In this article, we restrict our study to only the positive $\dot{\phi}_B.$ Then the value of $\phi_B$ is varied gradually and the goal is to find that value of $\phi_B$ which gives rise to $N_{inf}\simeq 60.$ The result for potential with linear monodromy term is tabulated in Table \ref{table1} and for a quadratic monodromy in Table \ref{table4}. It has been observed in Table \ref{table1} that more than $60$ number of e-folds have been already achieved with $\phi_B=0$ for linear monodromy case. And the value of $N_{inf}$ increases as we increase the initial value of the field $\phi_B$. While on the contrary, for potential dominated by quadratic monodromy term we obtain a minimum number of $60$ e-folds for $\phi_B=0.755$ Table \ref{table4}. Here we are interested in determining as to what value  of the ratio of kinetic to potential energy, $r_w= \frac{\dot{\phi_B}^2}{2}/V(\phi_B)$, at the bounce that would give rise to $\sim60$ number of e-folds. We call this value  to be the critical $r_w$ designated as $r_w^c.$ The value of $r_w^c$ for quadratic monodromy potential is found to be $r_w^c=1.28 \times 10^{12}.$

Next we consider the effects of variation of the parameters $b$, $f$ and $\delta$ on the background dynamics for both the potentials case by case. To this effect, we consider the variation of these parameters one by one on the value of $N_{inf}.$ To our observation, not shown here to avoid repetition, we found that, keeping the phase factor $\delta$ fixed, if we vary individually the parameters $b$ and $f$ there is hardly any noticeable change in the value of $N_{inf}.$ Therefore, we show here the effects of variation of the product $bf$ as a whole on the background dynamics whose effect is noticeable. The results for this are listed in Table \ref{table2} for linear and in Table \ref{table5} for potential with quadratic monodromy term. We note, for the linear case, Table \ref{table2}, as we increase the value of the product $bf$ there is a significant rise in the number of e-folds $N_{inf}.$ Thus, a greater number of e-folds of expansion can be achieved by tuning the parameters to higher value but less than one. In contrast, for potential with a square monodromy term, there is no significant change in the value of $N_{inf}$ with the variation of the product $bf$ shown Table \ref{table5}. This is because the quadratic term in the potential dominates heavily over the modulation term as compared to that of linear monodromy case. Thus the dynamics, by and large, is determined by the quadratic monodromy term rather than the modulation. The whole exercise is repeated for different initial conditions to arrive at the respective conclusion both for linear and quadratic monodromy term.

Finally, we study the effects of variation of the phase factor $\delta$ on the background dynamics. Here, also, we calculate $N_{inf}$ for different values of $\delta$ and repeat the same for four different initial conditions. In order to see the effect of $\delta$ we optimize the  product $bf$, which appears as a coefficient of the modulation term to $(b,f)=(0.9,0.9)$, giving $bf=0.81$. Since in the potential we have a cosine term, therefore, we set the range of our study to $\delta=[0,\pi]$ and divide it into four different values $\delta=(0, \frac{\pi}{4}, 2\frac{\pi}{4}, 3\frac{\pi}{4}, 4\frac{\pi}{4}).$ As shown in Table \ref{table3}, for linear monodromy case, the values of $\delta$ on $N_{inf}$ has, indeed, a drastic impact. For example, with $\phi_B=0$ the value of $N_{inf}$ is found to be $N_{inf} \approx 367$ for $\delta=0$, whereas, it is $N_{inf} \approx 33$ for $\delta=\frac{\pi}{2}.$ Therefore, while we obtain more than $60$ e-folds of expansion for one value of $\delta$ is not true for another value. And hence the value of $r_w^c$ would be significantly different for different values of $\delta$ in general. This observation can be understood to arise from the properties of the modulation term containing the sinusoidal function. On the contrary, in our analysis not shown here, we did not observe any significant change on the value of $N_{inf}$ upon the variation of $\delta$ for a quadratic monodromy potential case even if we set $(b,f)=(0.9,0.9).$ This again, we infer to happen because the square term is far more dominating in dictating the dynamics of the background than that of the modulation term. Hence, we conclude that in the quadric case parameters $\delta$ and $(bf)$ have negligible effects on the background dynamics as far as the calculation of $N_{inf}$ is concerned.
 
The final Sec.\ref{DySA} is dedicated towards a dynamical system analysis to draw qualitative  information of the universe for the two different monodromy potential considered in Sec.\ref{Cosmology} to study the background cosmology. This includes reformulation of the problem into two first order ordinary differential equations, for each case, by defining suitable pairs of dynamical variables. We adopt the standard formulation of linear stability analysis to comment on the stability of the fixed points if any and also draw qualitative and intuitive information by producing the phase portrait. Details of derivation towards the formulation of the dynamical systems are shown for both the potentials. For each potential we divide the section, further, into two parts one with the subheading ``phase portrait" and  the other ``fixed point analysis". Under the subheading ``phase portrait" we obtain the set of dynamical equations and comment on the qualitative behavior of the same. While the subsection under the ``fixed point analysis" deals in detail with rigour on the proofs of the statements made regarding the phase portrait. We notice that for the dynamical pair $(x,y)$ the phase space is not bounded for linear monodromy potential case. This is solely due to the profile of the potential. In this case of linear monodromy, the peripheri of the phase portrait is a slight deviation from the ellipse about x-axis with concave up towards negative x axis and its intercept lies on the positive x-axis. The slight deviation of the boundary from ellipse is due to the modulation term with small strength in the potential. On the other hand, for potential with quadratic monodromy term the periphery of the physical phase space is almost circular with a slight deviation due to the small modulation term. Significantly, in both the potentials, we note that trajectories gets attracted towards inflationary solution. It seems that the phase portrait for the linear monodromy case, the trajectories do not get settled down to a point after the inflation is over. This, pictorially, suggests the absence of any fixed point of the dynamical system governed by a linear monodromy term with a modulation.  And hence there is no finite equilibrium point in the phase space for the inflaton to settle down after the inflationary epoch is over. One needs to perform a global dynamical system analysis to comment on the ultimate fate of the phase space trajectories in this case which is beyond the scope of the present paper. This is followed by rigorous proof in subsection \ref{FiP1} to show that there, indeed, exists no fixed point, for linear monodromy case, with $b<1.$ On the other hand for quadratic monodromy potential case, the phase portrait clearly suggests the existence of a fixed point which is the origin of the phase space. The portrait, in this case, shows the trajectories spiraling around the origin $(0,0)$ before meeting. This is followed by rigorous proof in subsection \ref{FiP2} to show that $(0,0)$ is indeed a fixed point for allowed range of parameters pertaining to different values of $\mu$. Next, we examine the stability of the fixed point $(0,0)$ fixing $\mu=0.9$ which is used to generate the phase portrait. We find, after linearizing about the fixed point, the eigenvalues of the coefficient matrix of the perturbed equation are of the form $\lambda_{1,2}= w \pm \sqrt{D}$. Where $w<0$ and $D<0$ for the region of parameter space shown on the left of Fig.\ref{Region}. We  explicitly find that both for $\mu=0.9$ (used for producing the phase portrait) and $10^{-6}$ (used for cosmology of the background) the fixed point is stable with trajectories approaching towards it without any specific direction forming a spiral which is what we exactly obtain in our phase portrait. The only difference is that the allowed region of parameter space in $(b,f)$ in the case of $\mu=0.9$ is split into two regions. Whereas as we decrease the value of $\mu$ down to $10^{-6}$ the two regions spreads and overlaps to make the allowed region of parameter space $(0<b<1, 0<f<1)$. We also plot the value of $D$ the discriminant for  regions of parameters both for $\mu=0.9$ and $10^{-6}$ and show that $D$ always assumes negative value for the allowed range of parameters. It is only for the non allowed values of parameters that $D$ assumes positive sign.
In reference \cite{NewPaperB} one of the authors of the present work studies a fractional monodromy potential (with a modification which alleviates the problems of the monodromy potential \cite{43} in the reheating phase) in the framework of LQC, mLQC-II and mLQC-I. It is interesting to note that for all of these frameworks they have obtained the origin as a stable spiral in the post-bounce classical region for fractional monodromy potential with $p=2/3$ which is the result we obtain in our present work for $p=2$. But, we note that a direct comparison can not be drawn as our potential has a modulation term along with the monodromy.

Finally, we would like to note that in this paper   only the e-folds during the slow-roll inflation phase were calculated. Before this phase,  for the kinetic energy dominated case (which is the one studied in this paper), the expansion 
factor $a(t)$ is universal and given, for example, by Eq.(3.4) in 
\cite{Zhu01}, from which it can be read off the e-folds in the pre-inflationary phase, $N_c \equiv \ln(a_c/a_B) = \left(1+\gamma_B(t_c/t_{pl})^2\right)/6$, where $t_c$ denotes the moment that the equation of state of the inflaton vanishes $w_{\phi}(t_c) = 0$, and $\gamma_B \equiv 24\pi \rho_c/m_{pl}^4$. In Tables I-IV of \cite{Zhu01}, it was shown that $N_c \simeq 4 \sim 5$ with its exact values depending on the specific models considered. During this pre-inflationary phase, a very short period ($\dot{H} \ge 0$) of super-inflation was identified 
in \cite{Boj02,Singh06}, right after the quantum bounce. Since the expansion factor is given analytically by Eq.(3.4) during this short period \cite{Zhu01}, one can easily calculate the e-folds from the quantum bounce $t = t_B$ to the moment that the super-inflation is ended ($t = t_{H}$), at which we have  $\dot{H}(t_H) = 0$. Then, we find that $\delta{N}_c \equiv \ln(a_H/a_B) \simeq 0.111$.

\textbf{Acknowledgement}
A.W. is supported in part by the National Natural Science Foundation of China (NNSCF) with the Grants Nos. 11375153 and 11675145 and Q.W is supported by NSFC Grant No. 11675143. M. Sharma would like to acknowledge Dr. Tao Zhu for the valuable discussion.

\end{document}